\documentclass{iopjournal}

\usepackage{amsmath}
\usepackage{hyperref}
\usepackage{gensymb}
\usepackage{amsmath, booktabs, array}

\begin{document}
\bibliographystyle{iopart-num}

\articletype{paper} 

\title{Effect of Controlled Magnetic Island Bifurcation on Electron Diffusion}

\author{
Jessica Eskew$^{1,*}$\orcid{0000-0002-7020-5971},
D. M. Orlov$^{2}$,
B. Andrew$^{1}$,
E. Bursch$^{3}$,
M. Koepke$^{4}$,
F. Skiff$^{5}$,
M. E. Austin$^{6}$,
T. Cote$^{7}$,
C. Marini$^{2}$
and E. G. Kostadinova$^{1}$
}

\affil{$^{1}$Physics Department, Auburn University, Auburn, AL, United States}

\affil{$^{2}$Center for Energy Research, University of California San Diego, San Diego, CA, United States}

\affil{$^{3}$Department of Applied Physics, Columbia University, New York, NY, United States}

\affil{$^{4}$Department of Physics and Astronomy, West Virginia University, Morgantown, WV, United States}

\affil{$^{5}$Department of Physics and Astronomy, University of Iowa, Iowa City, IA, United States}

\affil{$^{6}$Institute for Fusion Studies, The University of Texas at Austin, Austin, TX, United States}

\affil{$^{7}$General Atomics, San Diego, CA, United States}

\affil{$^{*}$Author to whom any correspondence should be addressed.}

\email{jre0039@auburn.edu}


\begin{abstract}
Magnetic islands strongly influence cross-field electron transport in magnetized plasmas. In particular, bifurcations of the island topology modify the number and location of O-points, X-points, and separatrix boundaries, thereby altering diffusion pathways. In recent DIII-D experiments, external magnetic perturbations were used to rotate and periodically bifurcate the island on the q = 2 surface, causing a switchback between a q = 2/1-dominated structure and a narrower q = 4/2-dominated structure. To investigate how this topological change affects electron transport, we employ the field line tracing code TRIP3D with an implemented collisional operator. Thermal, tracer electrons launched from O-points, X-points, and outside separatrix boundaries reveal distinct diffusion regimes, including classical, subdiffusive, and superdiffusive behavior, depending on both the dominant island mode and launch location. These results suggest that island bifurcation can alter electron diffusion across rational surfaces, with direct implications for particle confinement. While the present work emphasizes diffusion as a general framework, the findings provide insight into the conditions under which electron trapping into an island or stochastization of the island's separatrix can enable additional mechanisms, such as the generation of energetic electrons.
\end{abstract}

\section{Introduction}
Magnetic islands are often formed in magnetized plasma due to dynamic processes such as reconnection and turbulence \cite{khab}. They are observed in space during solar flares, where the solar wind can drive reconnection in the Earth’s magnetosphere \cite{chen}, and in laboratory fusion devices such as tokamaks \cite{Kostadinova_eetransp, Textor1, Textor2}, stellarators \cite{killerw7x}, and reversed-field pinch plasmas \cite{clayton_mst}. The ubiquity of magnetic islands across plasma environments highlights their importance for transport processes: the width and structure of an island determine how electrons move along and across the magnetic field lines, which in turn affects confinement and loss.

In addition to influencing transport, changes in magnetic topology can contribute to processes that energize electrons. When the velocity of an electron exceeds a critical threshold, collisional drag is reduced and acceleration by electric fields can dominate, producing suprathermal or runaway populations that are effectively collisionless even in laboratory plasmas \cite{fleischmann_collisional_1992, heidbrink_mech}. Multiple mechanisms, including magnetic reconnection \cite{oieroset}, turbulence \cite{drake2003}, and interactions with magnetic islands themselves \cite{drake}, can lead to such energization. While electron transport in tokamak plasmas is also influenced by collisional and neoclassical processes and by turbulence, these effects are not investigated explicitly here. This work focuses on isolating how changes in magnetic field topology associated with island formation and bifurcation modify electron diffusion. Previous studies have shown that changes in island width and topology can modify the local distribution of electron energies \cite{boozer2016, chen}, suggesting a close interplay between energization and transport. For this reason, it is essential to understand how island topology alters electron diffusion at the fundamental level.

This paper investigates how magnetic island bifurcation affects the electron diffusion regime by analyzing data from recent Frontiers Science experiments, experiments that pursue a broad class of plasma physics research outside of strictly fusion-focused topics, at the DIII-D tokamak. In these experiments, static islands were grown and manipulated through 3D magnetic perturbations using internal perturbation coils (I-coils). The island rotation causes periodic bifurcation of the island structure located on the $q=2$ surface, which leads to a switchback from a homoclinic structure with two O-points and two X-points (dominated by the $q=2/1$ mode) to a heteroclinic structure with four O-points and four X-points (dominated by the $q=4/2$ mode).

Homoclinic islands are those with fixed X-points and O-points and connected flux surfaces. In contrast, heteroclinic islands are those with fixed X and O-points, as well as associated internal flux surfaces, which do not connect to other islands in a set \cite{evansbifur}. Previous studies have investigated magnetic island bifurcations in fusion devices \cite{wu_topological_2019}, \cite{bardoczi_bifur}, \cite{andreisobif}, \cite{hayashi_heliac}, and \cite{leal_isormp}; however, most of these studies have focused on heteroclinic bifurcations of tearing mode islands.

The present study focuses on the controlled growth and dynamics of static islands caused by resonant magnetic perturbations. The main question of interest is how changes in the island structure, such as bifurcations, affect electron trapping (or de-trapping) and cross-field diffusion. Specifically, we investigate how the bifurcation of the $q=2$ surface island affects electron diffusion in DIII-D. To examine how features of the island topology (width, number of O-points and X-points) affect electron diffusion, electron tracer simulations are employed by implementing a collisional operator in the magnetic field line tracing code TRIP3D \cite{evans_modeling_2002}.
TRIP3D with a simple collisional operator \cite{kalling_accelerating_2011} has been previously used to investigate electron diffusion in magnetic islands in NSTX-U simulations \cite{wu_topological_2019}. Here, we build on these previous studies by improving the physicality of the collisional operator and modeling recent DIII-D experiments. In the NSTX-U simulations, successive bifurcations were caused by increasing the magnitude of the perturbation coils current. As a result, the initial island chain transitioned into increasingly smaller island sub-substructures surrounded by large regions of field line stochasticity. In contrast, the island bifurcation studied here occurs as a periodic switchback between two distinct structures.

The main findings of the present study are summarized as follows. Electrons launched from different locations within the island topology are found to exhibit distinct diffusion regimes. Electrons launched at O-points are largely trapped, with stronger confinement observed in the larger 2/1 island, resulting in subdiffusive behavior. In contrast, X-points act as channels for transport, with the larger number and larger radial width of the X-points resulting in superdiffusion. The emergence of additional X-points in the bifurcated 4/2 case results in superdiffusive behavior as the bifurcation weakens confinement near X-points, making them preferential escape routes for radial spreading. Additionally, larger X-points in the 2/1 island are associated with more efficient transport. 

The remainder of this paper is organized as follows. Section \ref{sec:theory} provides theoretical background on the physics of electron diffusion and magnetic islands. Section \ref{sec:experiment} provides a brief overview of experimental results from the Frontiers Science DIII-D campaign, highlighting shot 196099. An overview of the TRIP3D code, collisional operator, and input parameters are provided in Section \ref{sec:trip3d}. In Section \ref{sec:results}, we show the TRIP3D results for shot 196099 and 10,000 particle tracers launched from three locations relative to the $q=2/1$ and bifurcated $q=4/2$ magnetic islands. In Section \ref{sec:quant}, we quantify the electron diffusion regime from histograms of particle tracer displacements. Island widths and Chirkov parameters are calculated using SURFMN \cite{Schaffer_surfmn} to further quantify the stochasticity of the magnetic field resulting from the island bifurcation. Section \ref{sec:discuss} provides a discussion of results, while Section \ref{sec:conclude} summarizes the main findings from the study. 

\section{Theoretical Background}\label{sec:theory}

\subsection{Island Formation and Topology}
In toroidal magnetically confined devices, like tokamaks, magnetic islands can result from perturbed magnetic fields due to (1) currents external to the plasma confinement region or (2) internal currents. Here, we investigate static islands caused by 3D magnetic perturbations from coil currents external to the plasma confinement region. These static islands are distinct from the tearing modes caused by internal currents, which can often spontaneously grow and disrupt the plasma \cite{boozer_islandsperturb}. 

An island in a magnetic field topology is a closed magnetic flux tube characterized by central elliptic points (O-point), hyperbolic points (X-points), and a separatrix surface, isolating it from the rest of space \cite{heidbrink_mech}. 
These island characteristics play a central role in the transport of energetic particles in magnetized plasma, as the interaction of the particles with different island features can result in different types of diffusive transport. Plasma particles at the separatrix follow hyperbolic trajectories. In contrast, particles within the separatrix are trapped in resonance and circulate the elliptic point (O-point) with a frequency proportional to the magnetic island width. 



Stochastic or chaotic regions of magnetic field lines can form when homoclinic tangles form near the island X-point \cite{park_plasma_2010}, \cite{roeder_homoclinic}, when neighboring island chains overlap \cite{biewer_MST}, or when a single island chain bifurcates into sub-island chains \cite{evansbifur}, \cite{bardoczi_bifur}. Stochasticity of the magnetic field lines can enhance cross-field transport due to chaotic particle trajectories \cite{horton1996chaos}. However, even for island overlap or bifurcation, the stochastic region coexists with remnants of magnetic islands, acting as attractors \cite{smith1978stochastic}. In these circumstances, nonlocal interactions are present in addition to chaotic trajectories, and the resulting particle transport can be subdiffusive, diffusive, or superdiffusive \cite{spizzo2018nonlocal}.

Magnetic islands in tokamak plasmas can exhibit internal bifurcations that result in the transition from homoclinic (single O-line) to heteroclinic island structure (multiple O-lines) \cite{bardoczi_bifur}. The O-line is the three-dimensional curve formed by connecting the O-points, which are the centers of closed magnetic flux surfaces within a magnetic island, along the toroidal direction. These bifurcations cause the emergence of additional islands with distinct O-points and independent separatrices. Here we consider a controlled transition from a homoclinic to a heteroclinic island caused by external coil perturbations. The initial structure is a homoclinic $q =2/1$ island chains: a single closed flux tube with a single O-line of $q =2/1$ helicity at the $q =2$ surface. During island rotations in our experiments, this homoclinic island was observed to transition to a heteroclinic $q =4/2$ island structure: two sets of $q =2/1$ islands residing at $q =2$, but their O-lines are disjoint.

\section{Experimental Results}\label{sec:experiment}

\subsection{Experimental Setup}

The DIII-D Frontiers Science experiments discussed here were used to study how changing topology of magnetic islands influences electron diffusion. These experiments built on findings from a previous study \cite{Kostadinova_eetransp}, which used an electron `tagging' technique (using electron cyclotron heating and current drive) to infer the presence of high-energy tail of the local electron distribution function at various locations within the plasma. These experiments showed that, for similar discharge conditions, tagging electrons inside versus outside the large island located at the $q=1$ surface resulted in qualitatively different energetic electron signature on data from emission diagnostics. Further, it was observed that suprathermal electrons are more pronounced when a larger region of stochastic magnetic field was created through island overlap in the edge plasma. The experiments analyzed here are from a follow-up campaign that aimed to understand the role of island chains in electron transport with and without stochasticity in the edge plasma. 
\begin{figure}
  \centering
  \includegraphics[width = \linewidth]{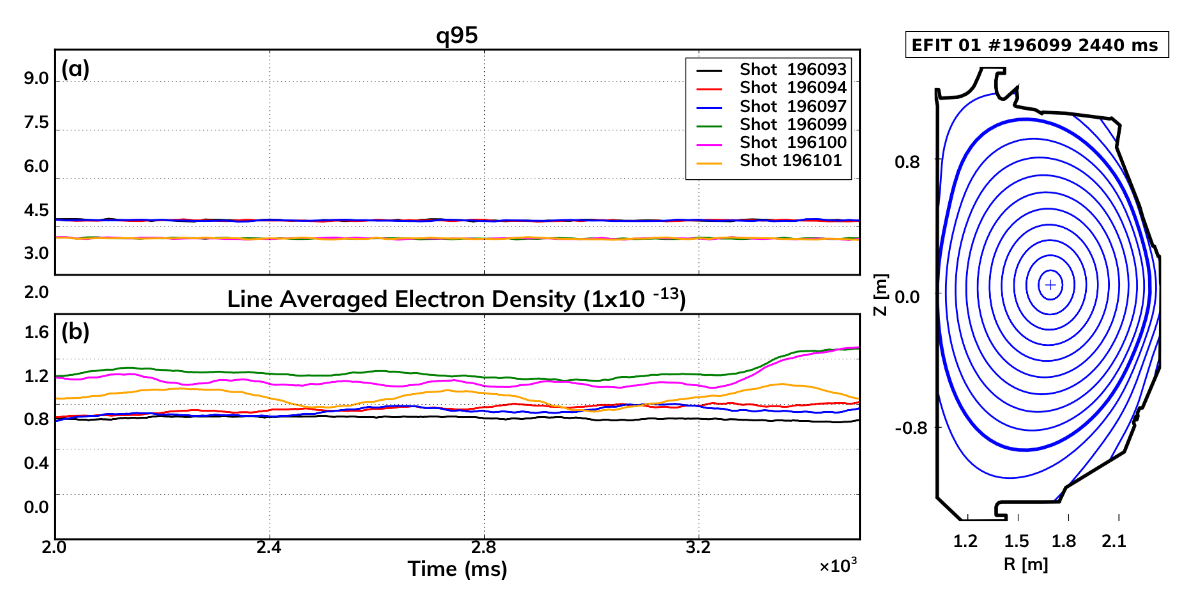}
  \caption{Plasma parameters (a) $q_{95}$ and (b) line averaged electron density for shots with magnetic island rotation during time interval 2000-3500 ms. Equilibrium reconstruction for shot 196099 at time slice 2440 ms.}
\label{fig:EFIT_prof}
\end{figure}

Each shot from the present campaign was an inner-wall limited L-mode discharge (fig. \ref{fig:EFIT_prof}) with no Neutral Beam Injection (NBI) heating, using only Ohmic heating and modulated Electron Cyclotron Heating (ECH)/Electron Cyclotron Current Drive (ECCD) pulse during flattop conditions ($I_P$ plateau). Beam blips at the end of each discharge were used to inform Charge Exchange Recombination (CER) and Motional Stark Effect (MSE) diagnostics. The toroidal magnetic field was kept constant at around $-2.10$ T. Plasma density was kept low $n_e = (0.9 -1.6) \times 10^{19} \ m^{-3}$ to avoid triggering sawtooth instabilities after $t = 1000 \ ms$. Such low density and simple geometry allowed for weak plasma response to magnetic perturbations, providing the ability to infer electron transport mechanisms from vacuum field simulations of the magnetic field using the field line tracing code TRIP3D (as discussed in Section \ref{sec:trip3d}). 

In this study, electron profiles were obtained using the Thomson Scattering (TS) system on DIII-D, which measures electron temperature and density over a range of 10 eV to 20 keV \cite{TSdiag}. These measurements provide the basis for determining plasma parameters in modeling during the formation and bifurcation of magnetic islands. Hard X-ray (HXR) \cite{heidbrink1986neutron} detectors were also monitored during the experiments. While the HXR system is primarily sensitive to high-energy electrons, its signals served as a useful reference: periodic peaks were observed to coincide with island bifurcation events, suggesting that bursts of energetic electrons were deconfined at a key phase of the rotation. This result is further discussed in Section \ref{subsec:obserEE}. Together, the TS measurements and the timing of HXR bursts establish the experimental context for investigating electron diffusion during island bifurcation.

\begin{figure}
  \centering
  \includegraphics[width = .7\linewidth]{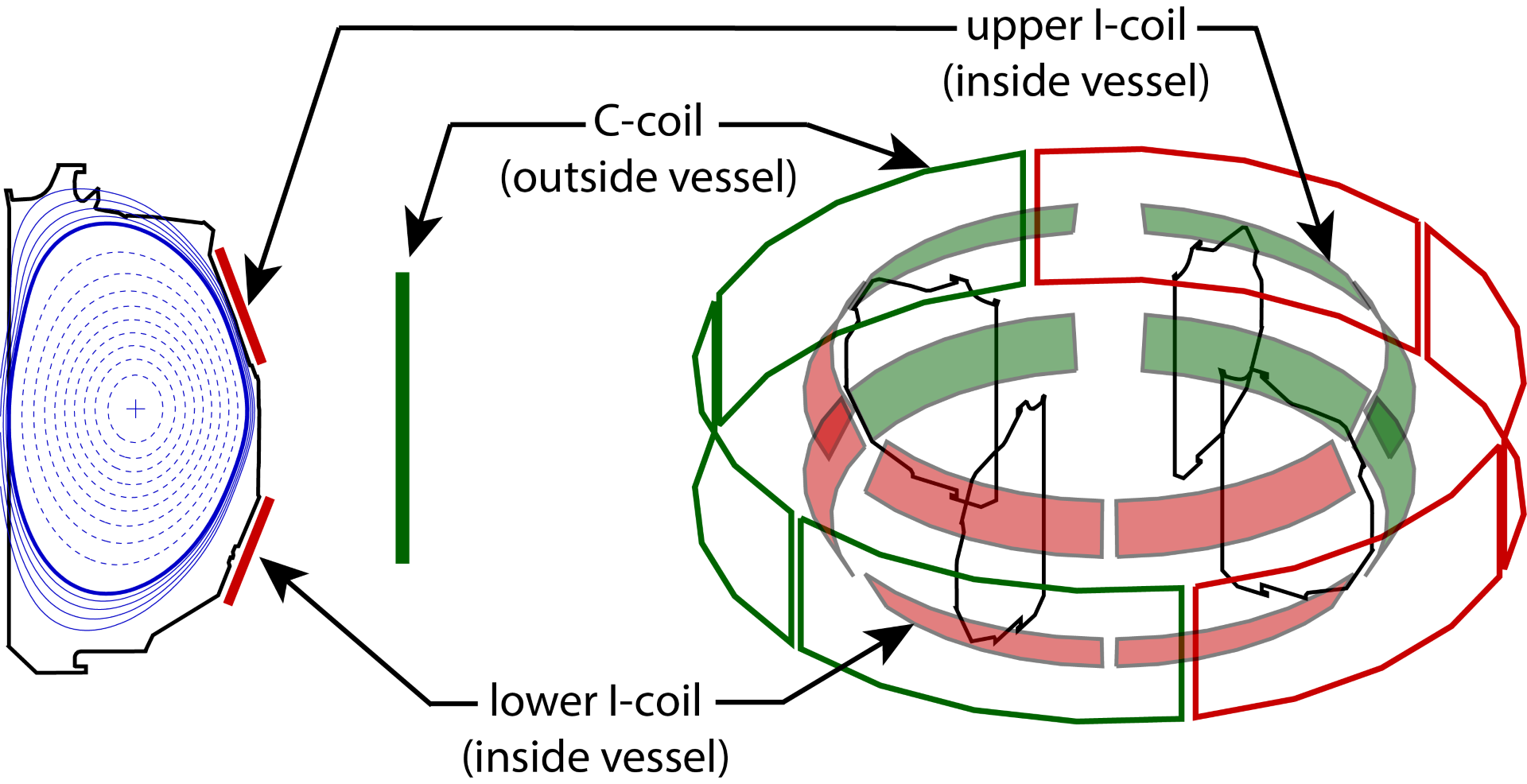}
  \caption{Diagram of coil geometry on DIII-D used for controlled formation and rotation of magnetic islands.}
\label{fig:COILS}
\end{figure}

Island formation in these discharges results from magnetic field perturbations from error field correction coils (outside vessel C-coils) and 3D Magnetic Perturbation (MP) coils (in-vessel I-coils), as shown in Figure \ref{fig:COILS}. In DIII-D, the in-vessel I-coil system consists of 12 window-pane coils, with six located above and six below the midplane and arrayed toroidally at 60° intervals, enabling the application of controlled non-axisymmetric radial magnetic perturbations \cite{schaffer_icoil}. Through appropriate phasing and current control of the upper and lower coil sets, this configuration permits the application of static and rotating $n=3$ fields and enables the generation and rigid rotation of an effective $n=1$ perturbation. Controlled island growth and manipulation were performed using the MP coils with $n=1$ perturbation, which results in island formation at all rational surfaces $q=m/n=m/1$. In some of the shots, the islands were rigidly rotated in a locked phase to the I-coil rotational frequency.

The present analysis focuses on cases where the islands were rotated by the I-coil with an amplitude of 2.5 kA, which produces a $\delta(Br)/B_{phi}~1e^{-4}$. In all discharges listed in Table \ref{tab:shotparam}, a bifurcation of the $q=2/1$ island into a $q=4/2$ island occurred at specific phases of the I-coil rotation. This bifurcation is attributed to destructive coupling between the rotating I-coil fields and the static perturbations from the C-coils and intrinsic error fields.

\begin{figure}
    \centering
    \includegraphics[width=\linewidth]{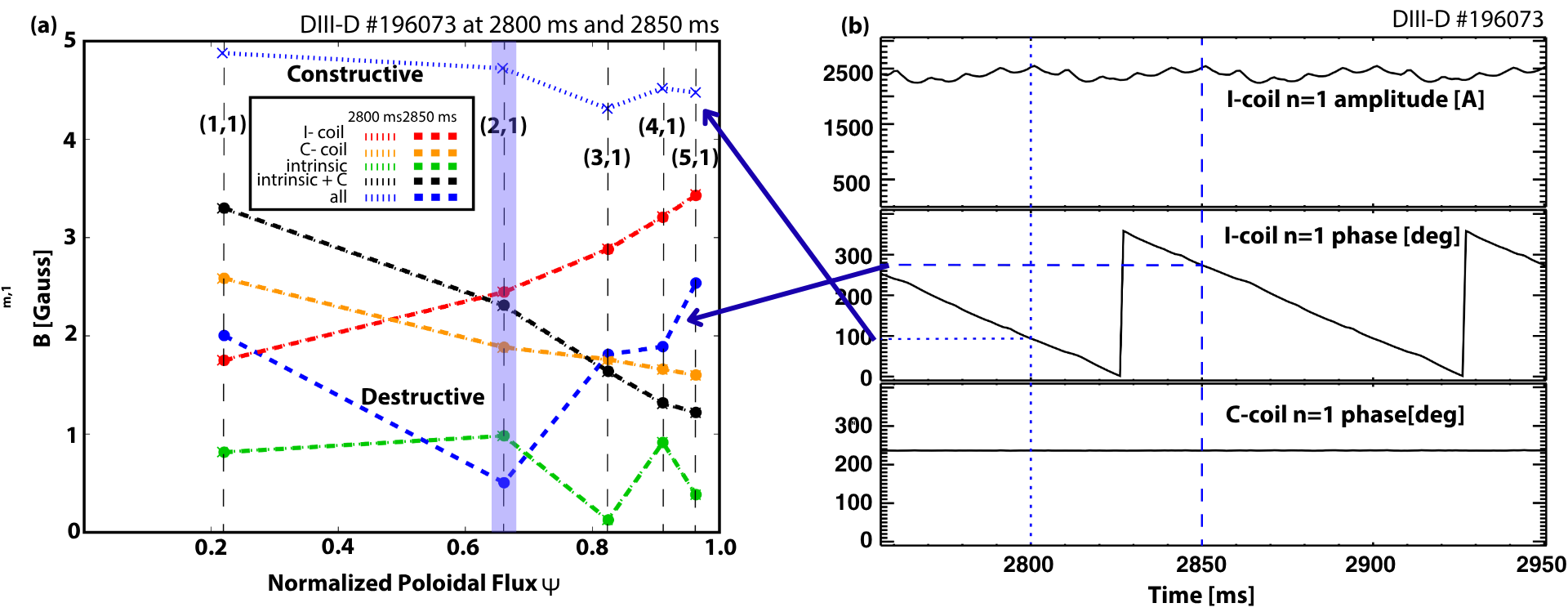}
    \caption{(a) Coupling between the rotating $n=1$ I-coil field (red dashed line) and fixed-phase perturbations from the intrinsic error field (dashed green line) and C-coil (dashed orange line). Radial profiles of the $n=1$ magnetic field spectrum ($B_{m,1}$) at 2800 ms and 2850 ms show how the total field at rational surfaces results from constructive or destructive superposition of the I-coil, C-coil, and intrinsic contributions. The $q=2/1$ surface (blue shaded region) is particularly sensitive to this phase relationship. (b) Time traces of I-coil amplitude, I-coil phase, and C-coil phase illustrate how the rotating I-coil periodically aligns constructively or destructively with the fixed C-coil and error fields, driving transitions in the island structure.}
    \label{fig:coilcouple}
\end{figure}

As shown in Figure \ref{fig:coilcouple}, both the I-coils and C-coils generate n=1 magnetic perturbations, but their effect on the plasma depends on their relative toroidal phase. When these coils are in phase with each other, their perturbations add together, doubling the effect inside the plasma (upper blue dotted line). Conversely, when the I-coil rotates 180° out of phase with the C-coil, their fields destructively interfere, nearly canceling the total n=1 perturbation at the $q=2/1$ surface (lower blue dashed line). This field cancellation eliminates the dominant island-driving harmonic, leaving only weaker sideband harmonics, which cannot sustain the original island. This precise phase manipulation leads to the observed bifurcation or destruction of the $q=2/1$ island and the emergence of a narrower $q=4/2$ island structure.

\begin{figure}
    \centering
    \includegraphics[width=\linewidth]
    {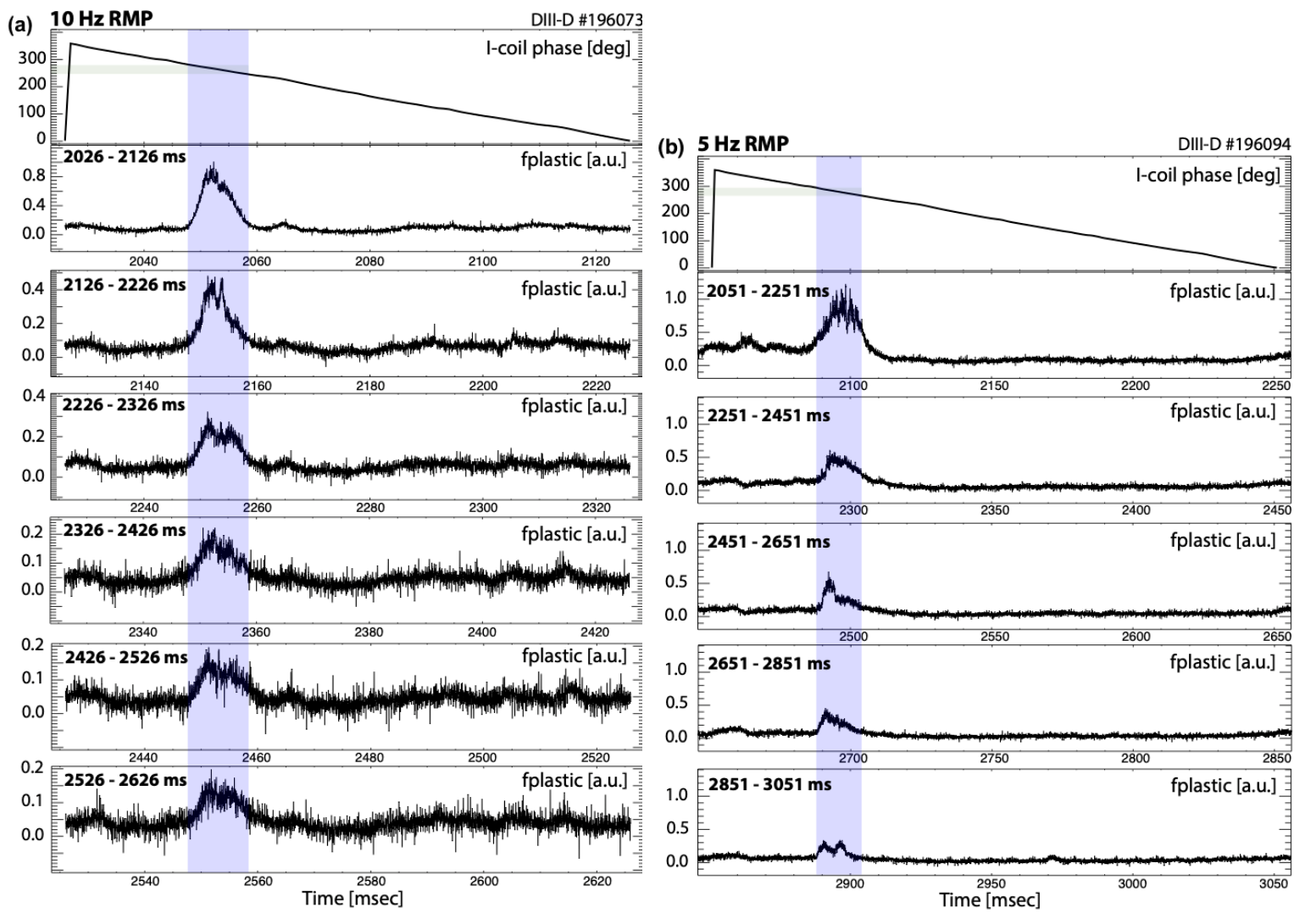}

    \caption{I-coil phase and Hard X-Ray (fplastic) for 10 Hz (a) and 5 Hz (b) rotational frequencies. Blue shaded regions indicate peaks in fplastic. Green shaded region shows consistent phase at which peaks occur in the I-coil phase.}
    \label{fig:phasefplast}
\end{figure}

Figure \ref{fig:phasefplast} shows how, at each peak from the Hard X-Ray detector and hence the times of bifurcation, the I-coil phase remains around 250-270$\degree$ roughly. This is consistent across shots with different rotation frequency. These plots show that regardless of how quickly the island is rotated, the bifurcation occurs at the same I-coil phase. 

Given the minimal plasma response and comparable density, magnetic field, and coil current amplitude in each shot, we will focus our attention on a single shot, shot 196099, selected as a representative for the remainder of the analysis. Observation and validation of this island bifurcation for shot 196099 is discussed below.

\begin{table}
  \begin{center}
\def~{\hphantom{0}}
  \begin{tabular}{lccccccc}
      $shot$  & $t[ms]$   &   Rot. Freq. $[Hz]$  &  $I_p [MA]$    & $q_{95}$ & $n [10^{19}m^{-3}]$\\[3pt]
       196093   &   6100    &   10      &   1.10    &   4.7    &   1.14 \\
       196094   &   6300    &   5     &   1.10    &   4.7   &   1.24 \\
       196097   &   5960    &   2   &   1.10    &   4.7    &   1.19 \\
       196099   &   4920    &   10    &   1.24    &   4.2 &   1.45 \\
       196100   &   4600    &   5     &   1.32    &   4.2   &   1.40 \\
       196101   &   4770    &   2    &   1.24    &   4.2   &   1.24 \\
  \end{tabular}
  \caption{Duration of discharge, I-Coil rotation frequency, plasma current, $q_{95}$, and average density for shots with island rotation.}
  \label{tab:shotparam}
  \end{center}
\end{table}

\subsection{Observation of Energetic Electrons and Island Bifurcation} \label{subsec:obserEE}

During island rotation, changing the phase of the I-coil current changes the contribution of the dominant wave mode responsible for the island on the $q=2$ surface. As a result, the structure transitions from a $q=2/1$-dominated into a $q=4/2$-dominated island chain, which is here referred to as "bifurcation". The $4/2$ structure has narrower island width and an increased number of O-points and X-points (from 2 to 4). This cycle repeats during successive island rotations, which results in a switchback between a structure dominated by the $q=2/1$ mode and a structure dominated by the $q=4/2$ mode. Interestingly, the timing of this island bifurcation coincides with the observation of peaks in Hard X-ray emission (HXR) diagnostics, suggesting periodic bursts of energetic electrons (EEs) escaping the plasma and hitting the wall.

\begin{figure}
  \centering   
  \includegraphics[width = .7\linewidth]{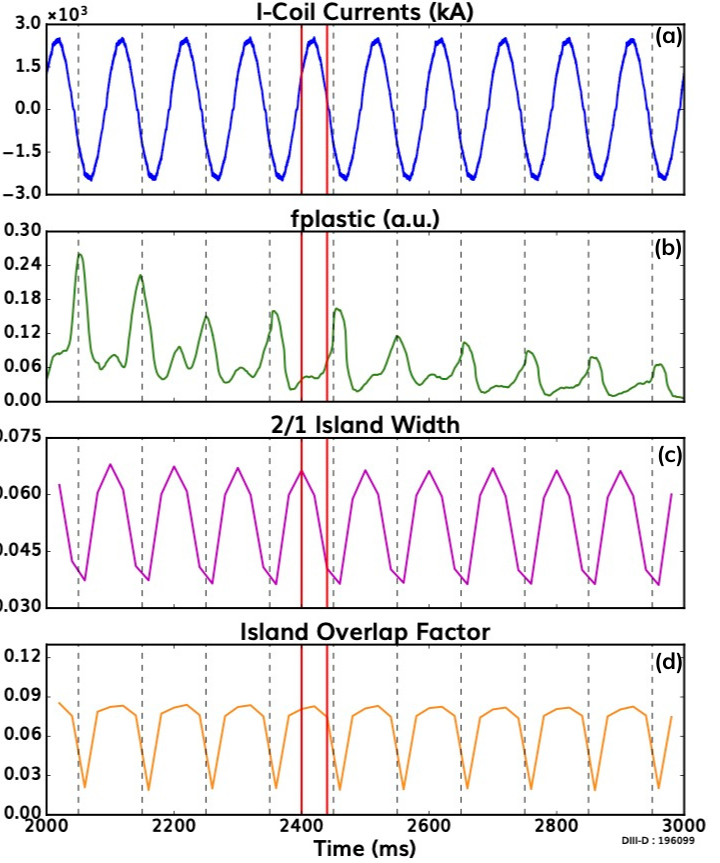}
  \caption{I-Coil current (a), X-ray scintillator detection of EEs (b), width of $q=2/1$ island (c), and vacuum island overlap width (d) for shot 196099 during the time interval 2000-3000 ms. Blue dashed lines represent RMP-induced island bifurcation, which occurs every 40-50 ms of the 10 Hz island rotation period and corresponds to peaks in X-ray data. Red lines show simulated times chosen (2400 ms and 2440 ms).}
\label{fig:196099_icfpwidthviow}
\end{figure}

Figure \ref{fig:196099_icfpwidthviow} shows the relationship between I-Coil current, HXR signal, $q=2/1$ island width, and the width of the vacuum island overlap in the edge plasma. The vacuum island overlap width represents the distance from the separatrix to the inner edge of the innermost overlapping magnetic island. This can be used to quantify chaotic or stochastic magnetic fields in the edge plasma. Dashed lines in Figure \ref{fig:196099_icfpwidthviow} represent the times of RMP-driven island bifurcation into a $q=4/2$ dominated structure, which occur every 40-50 ms for discharges with a 10 Hz island rotation frequency. One can see that, at times marked by dashed lines, the width of the $q=2/1$ island is at a minimum, while periodic bursts of EEs are detected as peaks in X-ray scintillators (fplastic) data. This happens consistently throughout the rotations, suggesting that the EEs were deconfined and hit the wall when the $q=2$ island bifurcated.  The amplitude of these bursts decreases with time, suggesting that the population of EEs diminishes after each island bifurcation and rotation cycle. 

It is unclear what physical process leads to releasing of EEs from islands. Previous work \cite{Kostadinova_eetransp} suggests that increasing the width of the stochastic magnetic field region in the edge plasma may be responsible for observing enhanced suprathermal electron emission. However, in the present experiments, the vacuum island overlap width in the edge reaches a minimum when the HXR peaks are observed. This result makes sense since it can be expected that when the wider $q=2/1$ island dominates, the edge magnetic surfaces are pushed closer together resulting in an increase in island overlap and stochasticity. In contrast, the collapse of the $q=2/1$ mode into a narrower $q=4/2$ structure leads to relaxed conditions on the edge islands and a decrease in the island overlap region. Thus, we conclude that the EE deconfinement observed here is likely due to processes occurring in the core plasma where the large RMP islands reside, and not due to stochasization of the edge plasma field. 

It is further unclear if the islands have a role in the electron acceleration and trapping. In space physics, both spacecraft data and numerical simulations have been used to propose that Fermi acceleration of electrons can occur within magnetic islands if the electrons become trapped and are repeatedly reflected by contracting magnetic field lines during island coalescence or reconnection \cite{chen}. In fusion research, it has been proposed that if electrons become trapped in islands during plasma disruption, they can experience acceleration to relativistic energies during the thermal quench \cite{boozer2016}. In both scenarios, it is important to understand what island topology would result in more efficient electron trapping.

Based on these DIII-D experimental results and mechanisms for electron acceleration discussed in literature, here we explore the hypothesis that topological changes of the magnetic field on the $q=2$ surface during I-coil-induced island rotation lead to a change in the electron diffusion regime, thus affecting the island's ability to trap or deconfine electrons. To help understand the physical mechanisms leading to the trapping and deconfinement, we focus on the investigation of electron cross-field diffusion as a function of changing magnetic field topology. Tracer electron simulations are conducted using the magnetic field line tracing code TRIP3D with a collisional operator to understand how changing topology during island bifurcation affects electron trapping and cross-field diffusion. 

\section{TRIP3D Code And Collisional Operator}\label{sec:trip3d} 
TRIP3D \cite{kalling_accelerating_2011}, \cite{evans_modeling_2002} is a field-line tracing code that computes engineering quality 3D magnetic field perturbation descriptions for most tokamaks performing or planning to perform 3D magnetic field and edge localized mode (ELM) suppression research (ITER, DIII-D, ASDEX-Upgrade, JET, MAST, NSTX, NSTX-U, KSTAR, EAST, TCABR, and SST-1). The TRIP3D suite offers the most complete available descriptions for all these devices, which serves as the starting point for much of the plasma response modeling performed to date. Specifically, TRIP3D integrates nonlinear magnetic field line differential equations (discussed in more detail below) and yields Poincaré maps of the poloidal magnetic field topology at a given toroidal angle. The code facilitates the exploration of magnetic island properties and stochastic field line topologies by determining the field line loss fraction. This fraction signifies the proportion of field lines intersecting a solid surface relative to each flux surface, with dependencies on toroidal revolutions and perturbation field amplitude.

TRIP3D is particularly useful for modeling experiments with small non-axisymmetric magnetic perturbations within a tokamak, which is the scenario of interest to the present study. The code efficiently integrates tens of thousands of field lines over several hundred toroidal revolutions, employing several thousand integration steps per revolution. Thus, TRIP3D allows for the construction of highly detailed Poincaré maps of the vacuum field topology, providing visualization of magnetic islands and regions of stochastic magnetic fields. In addition to obtaining the vacuum field topology, here we utilize TRIP3D as a particle tracer code through the implementation of a collisional operator. Particle tracer simulations enable the investigating how the characteristics of electron diffusion change with varying magnetic field topologies during island bifurcation. 

\subsection{Theoretical Overview of TRIP3D}

TRIP3D is used to trace magnetic field line trajectories using coil input parameters informed by experimental data and to construct Poincaré plots for analysis. The process starts by obtaining the unperturbed plasma equilibrium vector field $\vec{b_e}$ from the axisymmetric Grad-Shafranov equation:
\begin{equation}
    \Delta^* \psi = -\mu_0 R^2 p'(\psi) - \frac{1}{2} F(\psi)F'(\psi).
    \label{gradshav}
\end{equation}

Here $\psi$ is the poloidal magnetic flux per radian of the toroidal angle $\phi$ enclosed by a magnetic surface, $\mu_0$ is permeability of free space, $R$ is the major radius of the tokamak, $p'(\psi)$ is the pressure on the flux surface $\psi$, and $F=Rb_{\phi}$, where $b_{\phi}$ is the toroidal component of the magnetic field. Solving \eqref{gradshav} yields the toroidal ($b_{\phi}$) and poloidal ($b_R$, $b_Z$) components of the magnetic field, allowing for the construction of the resulting magnetic equilibrium vector field: \begin{equation}
\vec{b}e = b_R  \hat{R} + b_Z \hat{Z} + b_{\phi}  \hat{\phi}.
\end{equation}
The poloidal components are computed from the poloidal flux as:
\begin{equation}
b_R = -\frac{1}{R} \frac{\partial \psi}{\partial Z} ;
b_Z = \frac{1}{R} \frac{\partial \psi}{\partial R}.
\end{equation}


Eq. (\ref{gradshav}) is solved using the equilibrium fitting code EFIT \cite{lao_mhd_2005} as an input. EFIT provides accurate representations of the flux surfaces due to the axisymmetric external shaping coils (F-Coils). MHD effects are constrained by experimental magnetic and current profile data on DIII-D. Here, we used equilibrium reconstructions based on magnetic measurements as inputs for TRIP3D. Figure \ref{fig:EFIT_prof} shows the unperturbed plasma equilibrium for shot 196096 at 2020 ms. 

Using the plasma equilibrium and perturbation coil currents as input, TRIP3D traces the magnetic field lines by integrating the following set of 1$^{st}$ order cylindrical $(R,\phi,z)$ magnetic $(b_R, b_\phi, b_z)$ differential equations:

\begin{equation}
    \frac{\partial R}{\partial \phi} =  \frac{R b_R}{b_\phi} ; \frac{\partial z}{\partial \phi} =  \frac{R b_z}{b_\phi}, 
    \label{diffeqns}
\end{equation}

where $R$ is the major radius, $z$ is the vertical axis pointing in the direction of the torus axis of symmetry, $\phi$ is the toroidal angle, and $b_{\phi}$ represents the toroidal component of the magnetic field, while $b_R$ and $b_z$ represent the poloidal components. The integration is performed via a standard Runge-Kutta initial value algorithm with variable integration step size. A TRACER algorithm within TRIP3D manages the integration of individual lines and stores the $R, \phi, z$ position and $b_R, b_\phi, b_z$ magnetic data at each step along the toroidal integration path. Within this TRACER algorithm, we implemented a collisional operator that allows for using TRIP3D as a particle tracer code.

\subsection{Collisional Operator}
To assess electron trapping and diffusion across the bifurcating $q=2$ surface, we implemented a collisional operator in TRIP3D based on the formulation in \cite{kalling_accelerating_2011}. This operator introduces random perturbations to electron trajectories after they propagate a characteristic mean free path, thereby mimicking collisional scattering processes in a simplified manner. While not a complete kinetic treatment, this stochastic approach captures the essential effect of finite collisionality on electron diffusion in the presence of magnetic islands.

The random kick formulation is implemented within the TRIP3D tracer algorithm to approximate collisional diffusion of electrons across magnetic field structures. A statistically significant number of test electrons is followed, assuming that tracers are influenced only by the electromagnetic field and do not perturb the plasma equilibrium. Collisions may occur between the test electron and a background electron as it travels along a magnetic field line in the plasma, sending the test electron to another field line in its vicinity. The main inputs to vary in the random kick function are the characteristic length and the kick size. Here, the characteristic length represents the mean free path (as informed from experimentally-measured plasma conditions), and the kick size is selected to yield displacement of one Larmor radius or gyroradius. Once a tracer electron has traveled a distance greater than or equal to its mean free path, a kick is initiated of magnitude equal to its gyroradius multiplied by a random angle within 360${^\circ}$. Repeating this process produces a random walk across flux surfaces, allowing the model to capture diffusive transport processes associated with magnetic island structures.

Background electron density and temperature at the $q=2$ surface were obtained through core Thomson Scattering (TS) measurements. The electron density and temperature values are used to calculate the two collisional operator inputs as defined below. The mean free path $\lambda_{mfp}$ is defined as:
\begin{equation}
    \lambda_{mfp} = \tau_e v_{te},
\end{equation}
where $\tau_e$ is the electron collision time given by:
\begin{equation}
    \tau_e =\frac{1}{f_{p,e}} = (\frac{2 \pi \epsilon_0 m_e}{n_e e^2})^{\frac{1}{2}},
\end{equation}
and $v_{te}$ represents the thermal speed for electrons as given by:
\begin{equation}
    v_{te} = (\frac{2T_e}{m_e})^{\frac{1}{2}}.
\end{equation}

Here, $f_{p,e}$ represents the plasma frequency for electrons, $\epsilon_0$ is the permittivity of free space, $m_e$ is the mass of the electron, $n_e$ is the electron density, $T_e$ is the electron temperature, and $e$ is the electron charge.
The kick size is defined as the Larmor radius and given by
\begin{equation}
    \rho = \frac{m_e v_{te}}{e B},
\end{equation}

where $B$ is the magnetic field.

Plasma parameters for shot 196099 used for the collisional operator in TRIP3D are summarized in Table \ref{tab:pparam}. Comparable temperature and density profiles at the $q=2$ surface were observed across other discharges in the experimental campaign, ensuring that the collisional operator inputs chosen reflect the specific plasma conditions under which the bifurcation was observed.
\begin{table}
  \begin{center}
\def~{\hphantom{0}}
  \begin{tabular}{lcccccc}
      Time [ms]  & $v_{te}$ [m/s]   &   $\tau_e$ [s] & $\lambda_{mfp}$ [m] &  $\rho$ [m]\\[3pt]       2400   &   $8.75 \times 10^6$    &   $1.37 \times 10^{-5}$     &  $120.2 \times 10^3$  &  $2.47 \times 10^{-3}$ \\
       2440   &   $1.05 \times 10^7$    &   $2.27 \times 10^{-5}$    &  $242.8\times 10^3$ &   $2.98 \times 10^{-3}$\\
  \end{tabular}
  \caption{Plasma parameters for shot 196099 used for collisional operator in TRIP3D.}
  \label{tab:pparam}
  \end{center}
\end{table}

\section{Results}\label{sec:results}

\subsection{Island Bifurcation Observations with TRIP3D Vacuum Reconstructions}
To establish a baseline for comparisons with TRIP3D runs conducted using the collisional operator, we first present TRIP3D reconstructions of the underlying magnetic field topology without collisions. These simulations are used to identify features of the vacuum field magnetic topology, such as locations of increased stochasticity and island X-points, O-points, and separatrixes (the boundaries between the inside of the magnetic islands and the surrounding regions). Specifically, we are interested in TRIP3D vacuum field reconstructions that show the topological changes of the magnetic field during the bifurcation from a $q=2/1$ island to a $q=4/2$ island. While each TRIP3D simulation is time-independent, we can infer the evolution of the magnetic field structure by running the code for successive time slices during a bifurcation. Resolving the vacuum field topology for successive time slices confirmed that the periodic island bifurcation occurs at times consistent with the EE-induced peaks observed in the Hard X-ray data. Only the 2400 ms and 2440 ms cases are shown in this study, as they capture the primary structural transition associated with the bifurcation. 

\begin{figure}
  \centering   
  \includegraphics[width = \linewidth]{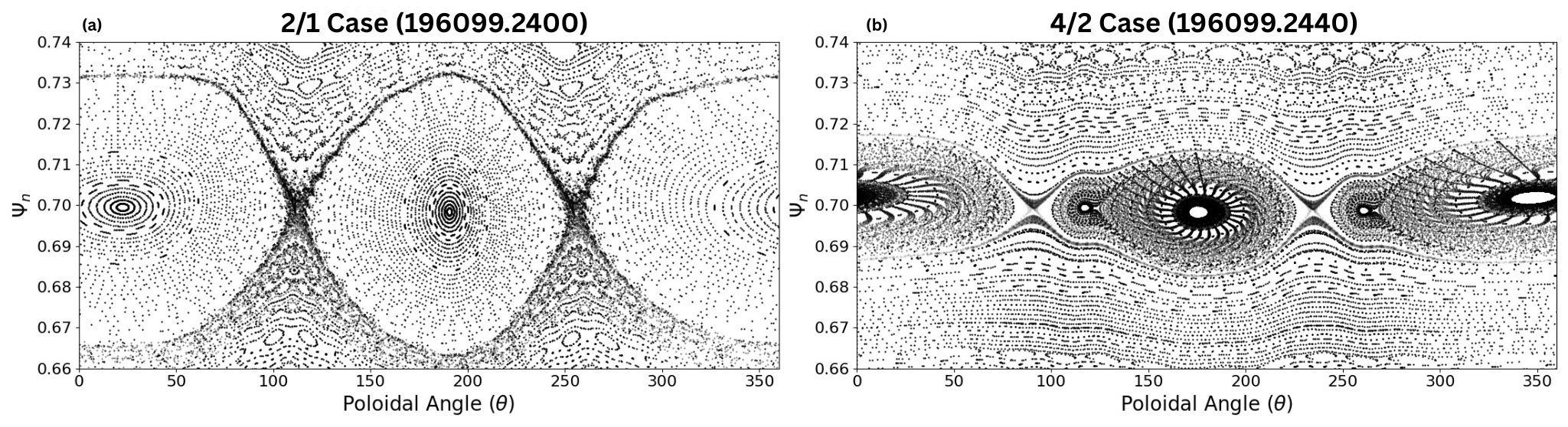}
  \caption{Vacuum field reconstruction of the $q=2/1$ island for shot 196099 at time slice (a) 2400 ms (showing the $q=2/1$ island) and (b) 2440 ms (showing the $q=4/2$ island).}
\label{fig:196099_vacuum}
\end{figure}

Figure \ref{fig:196099_vacuum} shows Poincare plots of the vacuum field reconstruction near the $q=2$ surface at a time when the $q=2/1$ mode dominates versus a time when the $q=4/2$ mode dominates the structure. The $q=2/1$ mode shows noticeable stochasticity within its envelope (near the X-points), while the bifurcated $q=4/2$ mode displays less, as indicated by the cleaner field line surfaces. A comparison of the two shows that, in addition to having 4 O-points and 4 X-points, the $q=4/2$ structure is visibly narrower. While the envelope width evolves due to the applied changing 3D perturbations, the appearance of multiple narrower sub-islands in the bifurcated $q=4/2$ mode may have important implications for confinement. As critical island width is needed to confine electrons with certain energy \cite{boozer2016}, the reduction in island width for the $q=4/2$ structure suggests that the observed spikes in the HXR detectors may be due to deconfinement of EEs with energies exceeding the threshold that the $q=4/2$ island can confine. This motivates the hypothesis that the observed spikes in HXR detectors result from the deconfinement of such electrons as the magnetic topology bifurcates. To assess how this change in the field structure affects electron diffusion, tracer electrons are launched from various locations relative to the island structure for both time slices.

\begin{figure}
    \centering
    \includegraphics[width=\linewidth]{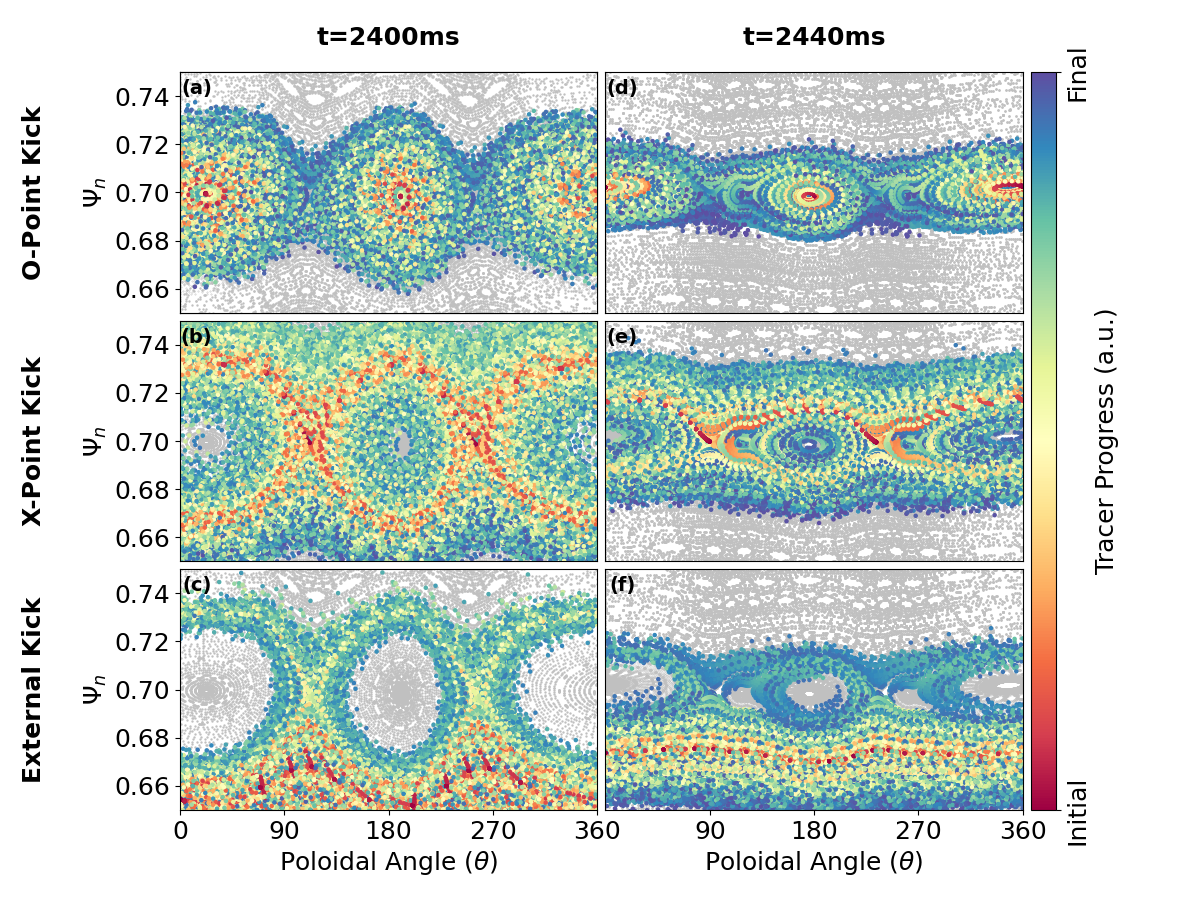}
    \caption{Results for 10,000 tracer electrons launched from different points in the two bifurcation phases. Left column (a-c) shows tracer diffusion across the $q=2/1$ island for shot 196099 at time slice 2400 ms. Right column (d-f) shows tracer diffusion across the $q=4/2$ island for shot 196099 at time slice 2440 ms. In each case, tracers were launched from an O-point (a,d), a X-point (b,e), or a point outside the separatrix line (c,f). Here color stands for increasing simulation step, i.e., earlier positions are in red, while later positions are in purple}
    \label{fig:196099results}
\end{figure}

\subsection{Comparison of Electron Diffusion across Different Island Structures}
Tracer electrons were launched from different locations relative to the island structure for the $2/1$ and the $4/2$ case. The electron trajectories were evolved under the action of the perturbation coil currents and the collisional operator in TRIP3D. Figure \ref{fig:196099results} shows the results from TRIP3D simulations with collisions for discharge 196099 at time slices before (2400 ms) and after (2440 ms) the island bifurcates from a $2/1$-dominated to a $4/2$-dominated structure. Ten thousand tracer electrons are launched from the O-points (fig. \ref{fig:196099results} a,d), X-points (fig. \ref{fig:196099results} b,e), and outside the separatrix surfaces (fig. \ref{fig:196099results} c,f). In these figures, color corresponds to time steps in the simulation, with red representing earlier times and purple representing later times. In this section, we utilize color mixing as a visual aid to enhance understanding of the tracer electron trajectories and qualitative assessment of the corresponding diffusion. In the next section, we will quantify the diffusion using non-Gaussian fits to the histograms of the tracer displacements. If the cross-field diffusion is classical, one expects the color of the tracer trajectories to gradually change from red to blue as the tracers move radially away from their launching point. However, in the presence of trapping effects or nonlocal jumps (big displacements) resulting in anomalous diffusion, one expects to see color mixing throughout the simulation region.

When launching from the O-point at the time when the $2/1$ mode dominates the structure (fig. \ref{fig:196099results} a), tracer electrons are seen to explore the entire island with a clear radial spread; however, they remain largely confined within the island's separatrix. Color mixing between red and purple is observed throughout the island, indicating that even early on, the tracers could make big jumps across the island's structure, followed by jumps backwards, which kept them within the separatrix. This color mixing can be explained by the observed stochasticity within the $2/1$ island envelope. In the presence of stochastic or chaotic fields, the electrons can make bigger displacements, connecting different regions of the island. However, there is also a high probability for the electrons to jump backwards toward the attractor O-point. Thus, we expect that in this case, the electron diffusion will deviate from classical and exhibit both forward and backward nonlocal jumps in displacement. In contrast, when the structure is dominated by the $4/2$ mode (fig. \ref{fig:196099results} d), tracer electrons diffuse radially outward with minimal color mixing (i.e., red colors are observed mostly closer to the launching point, while blue colors are observed mostly around the separatrix. Since the $4/2$-dominated structure did not exhibit stochasticity inside the island separatrix, we can expect that the electron diffusion in this case is close to classical or mildly subdiffusive due to the presence of the attractor O-point.

Similar trends in color mixing are observed when the tracers are launched from X-points. Tracer electrons launched from the X-point of the $2/1$ structure (fig. \ref{fig:196099results} b) exhibit increased outward diffusion with the tracers exploring a larger radial range when compared to tracers launched from the X-point of the $4/2$ structure (fig. \ref{fig:196099results} e). In the $2/1$ case, tracer electrons can penetrate the island but also diffuse both inward and outward, as evident from the color mixing (i.e., we see a mix of red and purple points both inside and outside the separatrix). When launched from the X-point of the $4/2$ structure  (fig. \ref{fig:196099results} e), the tracer electrons exhibit smaller radial spread both inward and outward of the island separatrix. While the color mixing for the $4/2$ case is visibly less than that for the $2/1$ structure, narrow regions of mixed red and purple points are observed around the island separatrix. This result suggests that the $4/2$ structure may also lead to deviations from classical electron diffusion for electrons interacting with the X-points. 

Launching from outside the separatrix for the $2/1$ dominated structure (fig. \ref{fig:196099results} c) shows that as the tracer electrons diffuse radially outwards, many diffuse through the  $q=2/1$ island separatrix. However, the tracers mainly interact with the X-point and do not seem to penetrate deep into the $2/1$ island. It can also be seen that the color mixing in this case is less pronounced as compared to Figure \ref{fig:196099results} a) and b), with most of the red color remaining around surfaces close to the launch location, while most of the purple color ends up around the separatrix. Similarly to previous cases, when launching from outside the $4/2$ island separatrix (fig. \ref{fig:196099results} f), the tracer trajectories show clear demixing of colors with red colors seen close to the launching locations. In contrast, purple colors are seen radially outward. However, in the $4/2$ case, the electrons also seem to penetrate the island more easily than in the $2/1$ case. Finally, for the same number of tracers, launch locations, and simulation time steps, we observe that the electrons interacting with the $2/1$ island are transported to larger $\psi_n$ than those interacting with the $4/2$ structure. This can be attributed to the larger X-points in the $2/1$ case and the associated stochasticity of surfaces surrounding the separatrix. 

\section{Quantitative Characterization of Tracer Diffusion and Stochasticity}\label{sec:quant}

To quantify the transport visualized in Figure \ref{fig:196099results}, histograms of the final locations and poloidal angles of tracer electrons were constructed and fitted to non-Gaussian distributions to assess tailness and deviations from classical diffusion. Additionally, we construct histograms of magnetic field line displacements from their original positions and quantify field line stochasticity using the Chirikov parameter. 

\subsection{Quantifying Tracer Diffusion}

\begin{figure}
    \centering
    \includegraphics[width=\linewidth]{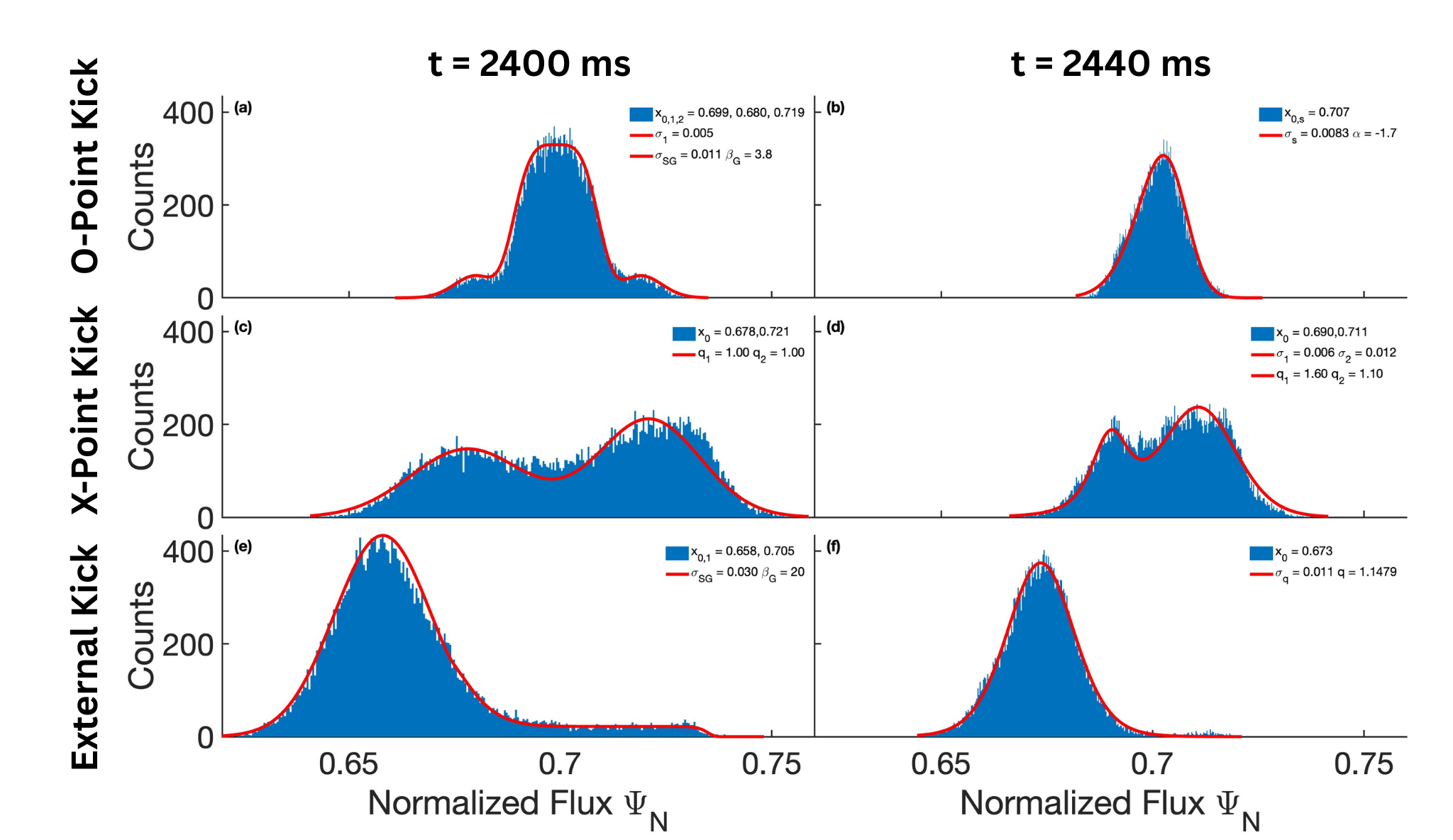}
    \caption{Final position histograms for each kick location: O-point ({a,d}), X-point ({b,e}), and location outside the separatrix ({c,f}). Left column ({a–c}) corresponds to the 2/1-dominated structure at time slice 2400 ms, while the right column ({d–f}) corresponds to the 4/2-dominated structure at time slice 2440 ms.}
    \label{fig:tripfinalhist}
\end{figure}

As visually represented in Figure \ref{fig:196099results}, the cross-field diffusion of the tracer electrons depends on their initial launching location relative to the topological features of the island structures. Figure \ref{fig:tripfinalhist} shows the histograms of final tracer positions launched from different locations (O-points, X-points, and outside the separatrix) for the $2/1$ and the $4/2$ island structure. In each case, distributions were fitted to the histogram to assess deviations from classical diffusion (i.e., deviations from Gaussian distribution of particle displacements). The various distribution functions used are described below. Common variables used include $\sigma$ representing the variance,  $\mu$ representing the mean, and $x$ as the input variable (i.e. the tracer's position).

A standard Gaussian distribution was used to model classical diffusion, which would be expected if electron diffusion is not appreciably affected by the underlying island topology (i.e., when the tracer motion is mostly due to the random collisions) 

\begin{equation}
    p(x) = \dfrac{1}{\sigma \sqrt{2\pi}} \exp\left( -\dfrac{(x - \mu)^2}{2\sigma^2} \right).
    \label{eqn:gauss}
\end{equation}

To account for asymmetric diffusive transport, such as that resulting from the emergence of a smaller sub-island in the $4/2$ case, a skew normal distribution was used

\begin{equation}
    p(x) = \left(1 + \operatorname{erf}\left( \dfrac{\alpha (x - \mu)}{\sqrt{2} \sigma} \right) \right) \exp\left( -\dfrac{(x - \mu)^2}{2\sigma^2} \right).
    \label{eqn:skewnorm}
\end{equation}

Here, skewness is quantified by the variable $\alpha$. A negative value of $\alpha$ indicates a negative skewness with the peak of the distribution pulled to the right and broader left tail. 

In cases exhibiting anomalous diffusion in the form of non-Gaussian tails, a generalized normal distribution was used to provide the best fit
\begin{equation}
    p(x) = \dfrac{\beta}{2\sigma \Gamma(1/\beta)} \exp\left( -\left|\dfrac{x - \mu}{\sigma}\right|^{\beta} \right).
    \label{eqn:gennorm}
\end{equation}

Here, the variable $\beta$ quantifies the deviation from classical diffusion. For $\beta > 2$, the distribution shows truncated tails, indicating subdiffusion. For $\beta < 2$, the distribution shows fat tails, indicating superdiffusion. When $\beta = 2$, the distribution limits to the normal Gaussian indicating classical diffusion. 

Finally, for several cases where the distributions exhibited "fat" tails, suggesting superdiffusion, we used the $q$-Gaussian distribution

\begin{equation}
    p(x) = \dfrac{A_q}{\sqrt{\pi \sigma^2}} \left(1 + \dfrac{(1 - q)(x - \mu)^2}{\sigma^2} \right)^{-\frac{1}{1 - q}}.
    \label{eqn:qgauss}
\end{equation}

Similar to $\beta$, the variable $q$ quantifies how the tails of the distribution deviate from a Gaussian distribution. For $q>1$, the q-Gaussian has heavy tails, indicative of superdiffusion or large jumps, and even Lévy flights for $q>5/3$. This suggests enhanced diffusive transport, where some particles move much farther than expected under normal diffusion. When $q<1$, subdiffusion is expected, suggesting slower-than-classical diffusion. 
Each distribution was selected based on its ability to reproduce the observed statistical spread of tracer final positions and to provide physical insight into the underlying diffusion regime.

In the case of the tracers launched from the $2/1$ O-point before bifurcation, shown in Figure \ref{fig:tripfinalhist} a), we see three peaks in the distribution, Two of them are small Gaussian distributions, corresponding to normally diffusive populations around the inner region of the $2/1$ island separatrix (with distribution peaks centered at $\psi_N = 0.68$ and  $\psi_N = 0.719$). A third large distribution is observed to be centered at a radial location of $\psi_N = 0.699$, which coincides with the location of the O-point. This distribution suggests subdiffusion, as indicative of the $\beta$ value greater than 2.  These results suggest that for the electrons launched from the $2/1$ island O-point, two distinct diffusion regimes are expected: subdiffusion for particles inside the separatrix near the O-point and classical diffusion for particles approaching the separatrix and the X-points. For the $4/2$ case after the bifurcation (fig. \ref{fig:tripfinalhist} d), the resulting final positions of tracer electrons launched from the O-point are best described by a negatively skewed (skewed normal) distribution, with the variable $\alpha = -1.7$, suggesting a moderate to strong negative skew and a long left tail. This skewness can be explained by the smaller sub-island emerging from the $4/2$ mode. Symmetry is broken due to the emergence of this substructure, indicating a preferred direction of the diffusion in this case. 

When launched from the $2/1$ island X-point before bifurcation (fig. \ref{fig:tripfinalhist} b), two Gaussian distributions emerge around each of the island's X-points. This is in agreement with the two Gaussian distributions observed around the separatrix regions in the $2/1$ case for tracers launched from O-points (fig. \ref{fig:tripfinalhist} a). For the $4/2$ case after the bifurcation (fig. \ref{fig:tripfinalhist} e), the tracer electrons are observed to remain around the X-points of the $4/2$ structure, similar to the case before the bifurcation. However, both distributions are non-Gaussian (with $q>1$ for both peaks), suggesting superdiffusion. The asymmetric feature of the diffusion observed for tracers launched from the O-point in the $4/2$ case (fig. \ref{fig:tripfinalhist} b) is confirmed for tracers launched from the X-point (fig. \ref{fig:tripfinalhist} d), with $q=1.6$ for one distribution and $q=1.1$ for the other.

For tracer electrons launched outside the $2/1$ island separatrix before the bifurcation (fig.~\ref{fig:tripfinalhist} c), the resulting final positions form a Gaussian distribution indicative of classical diffusion, centered around the inner separatrix boundary of the $q=2/1$ island ($\psi_N = 0.658$). A smaller, population of particles forms a "plateau" tail that extends radially outward through $\psi_N = 0.74$, which coincides with the radial extent of the X-points. This result supports the interpretation that, while the $2/1$ island structure can act as a transport barrier for particles near the O-point, the extended X-point regions of the $q=2/1$ island serve as preferential pathways for radial excursions, allowing efficient transport for a small fraction of particles across the island. For tracers launched outside the $4/2$ island separatrix after the bifurcation (fig.~\ref{fig:tripfinalhist} f), the distribution of final tracer positions is best described by a q-Gaussian distribution, exhibiting symmetric superdiffusion with $q=1.15$. When compared to the $2/1$ case (fig.~\ref{fig:tripfinalhist} e), the radial displacement of the tracers is overall smaller reaching  $\psi_N = 0.71$ for the same initial launch location and same number of time steps. This suggests that the radial extent of island X-points plays a key role in the efficiency of electron transport across the island.

\begin{figure}
  \centering   
  \includegraphics[width = .45\linewidth]{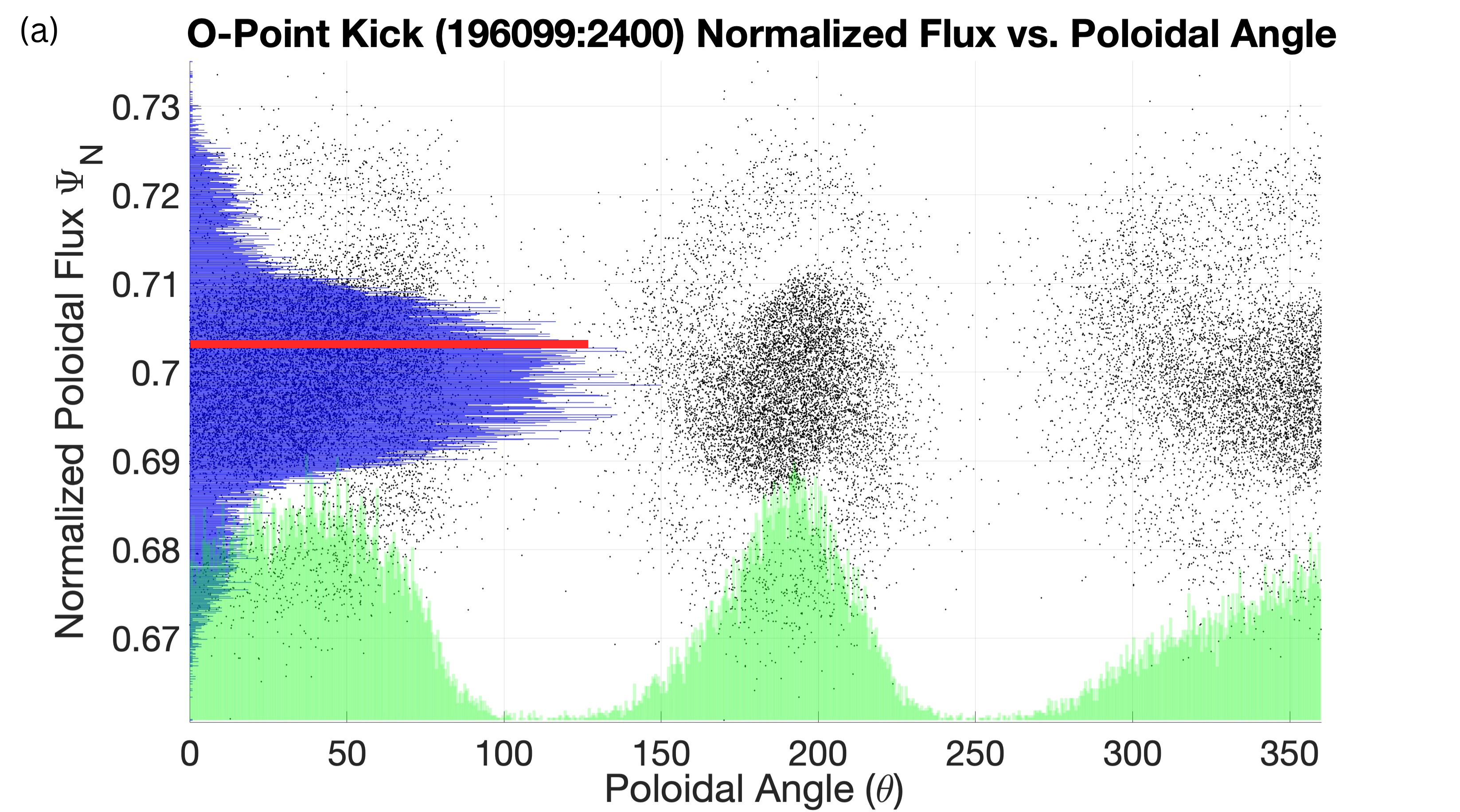}
  \includegraphics[width = .45\linewidth]{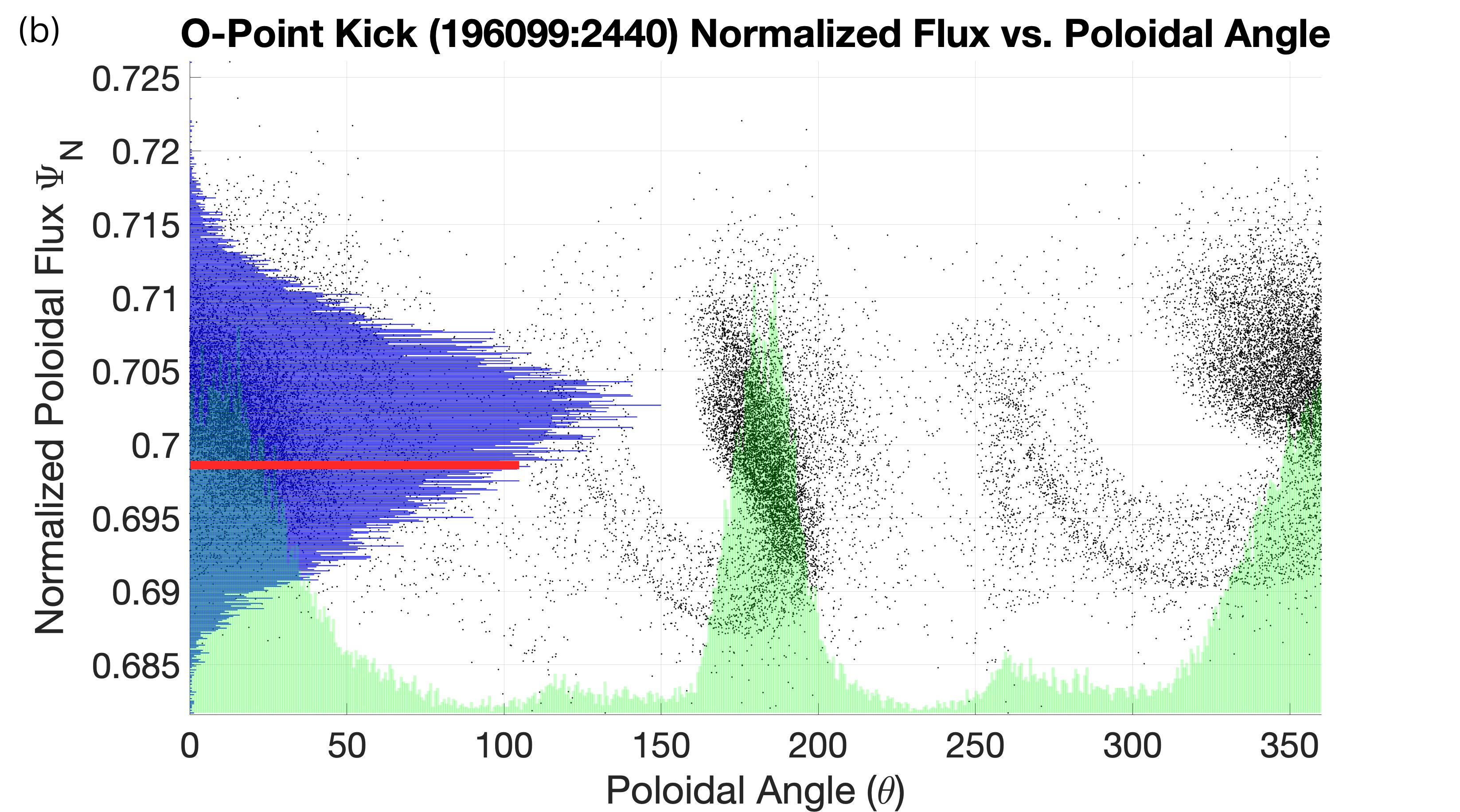}
  \includegraphics[width = .45\linewidth]{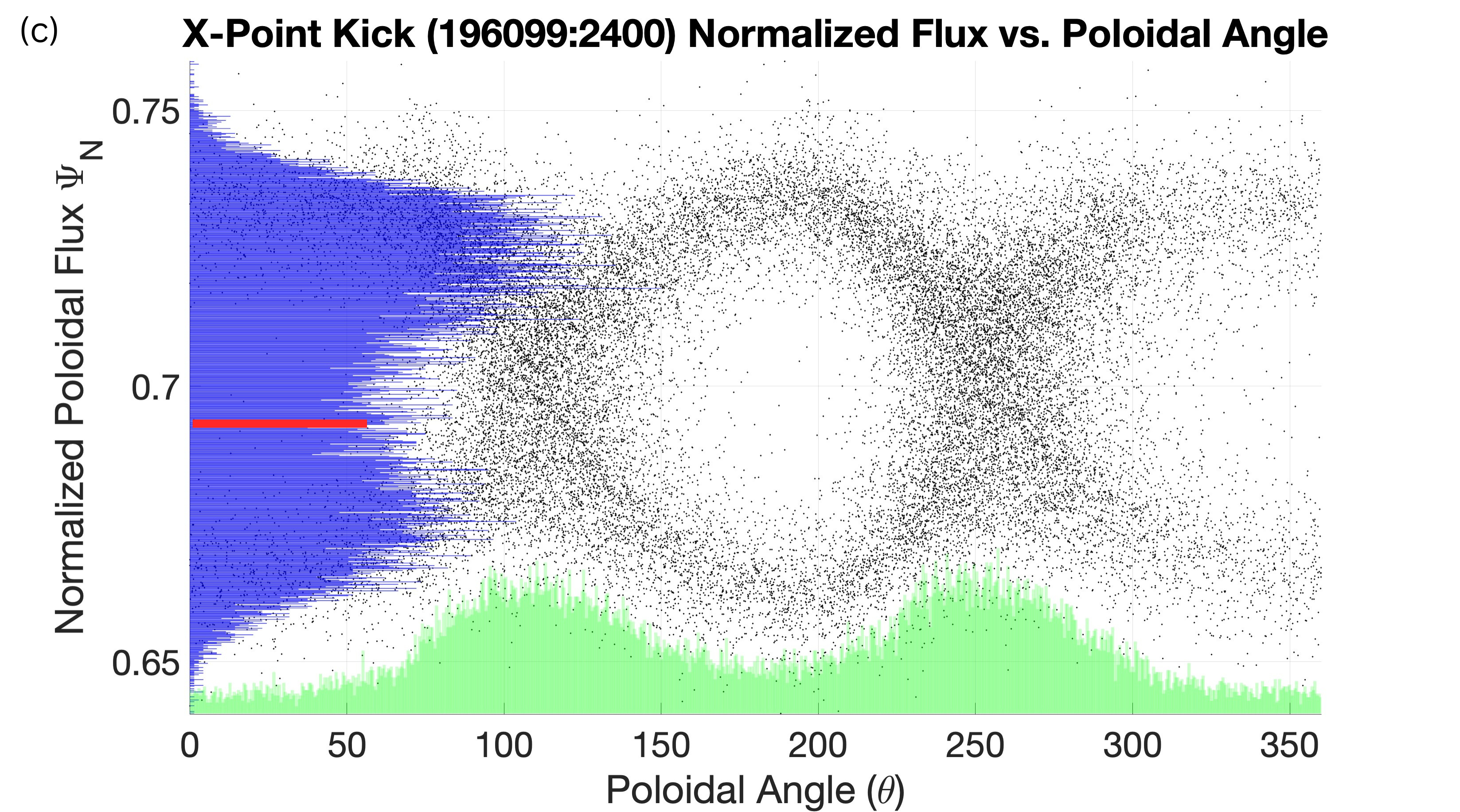}
  \includegraphics[width = .45\linewidth]{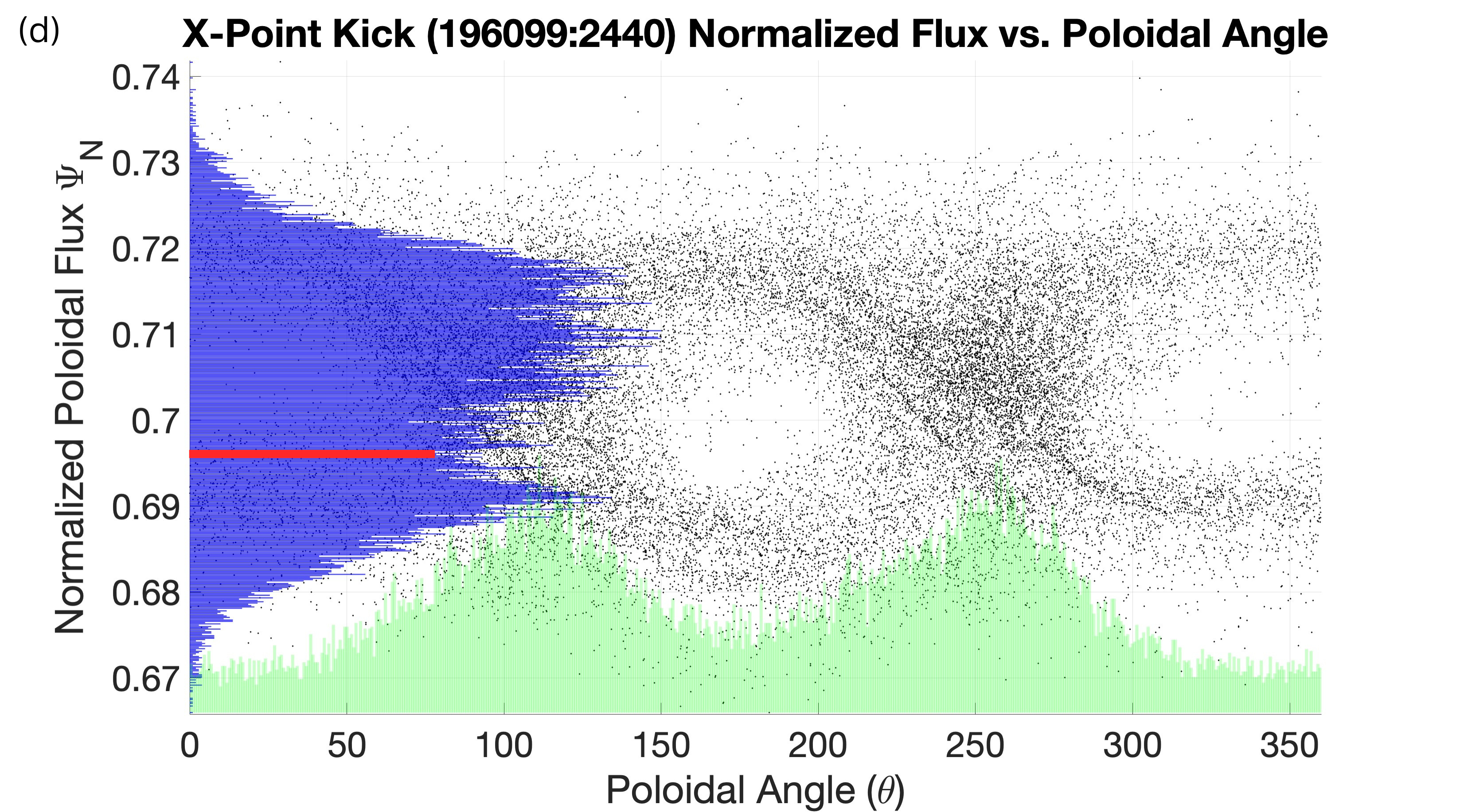}
  \includegraphics[width = .45\linewidth]{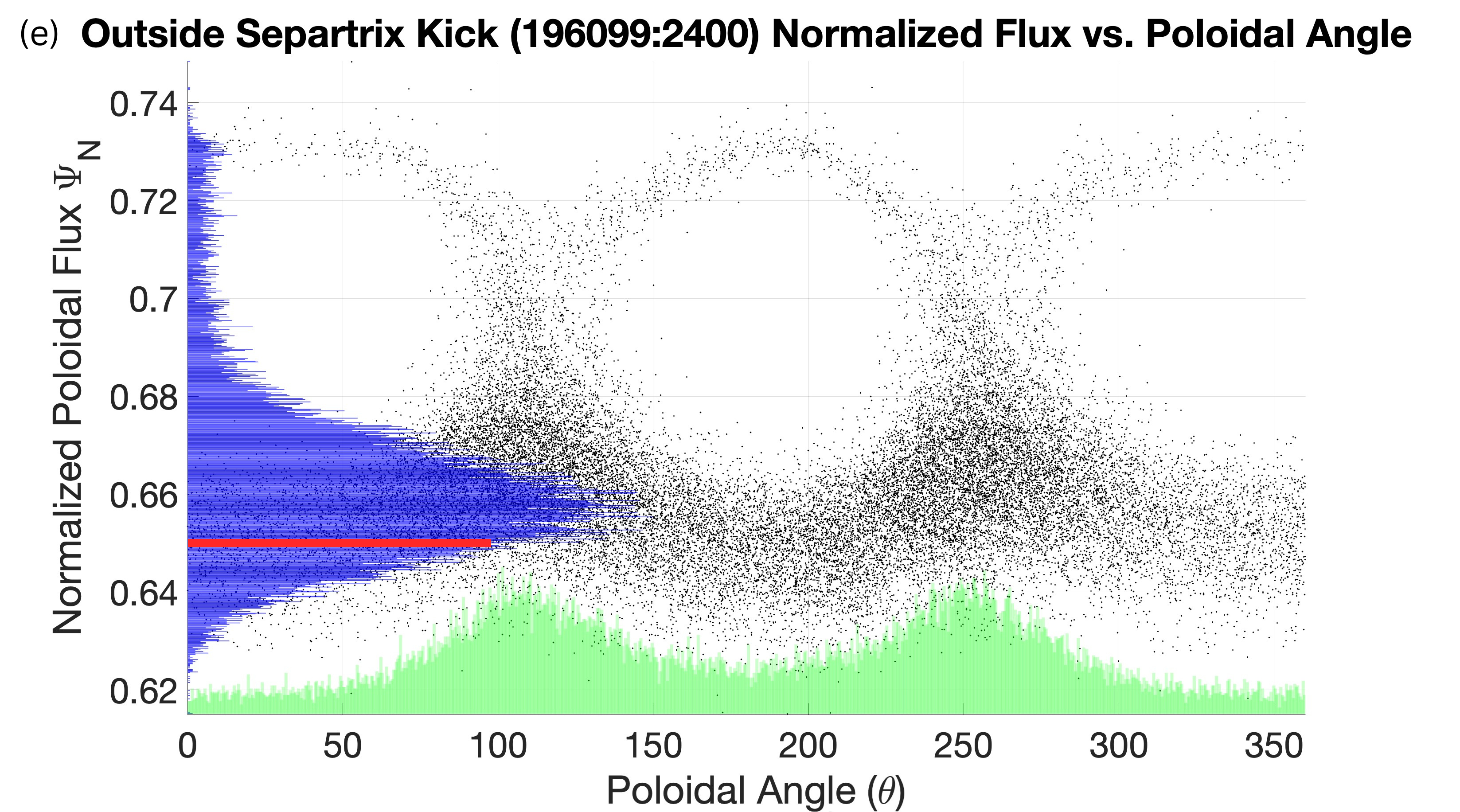}
  \includegraphics[width = .45\linewidth]{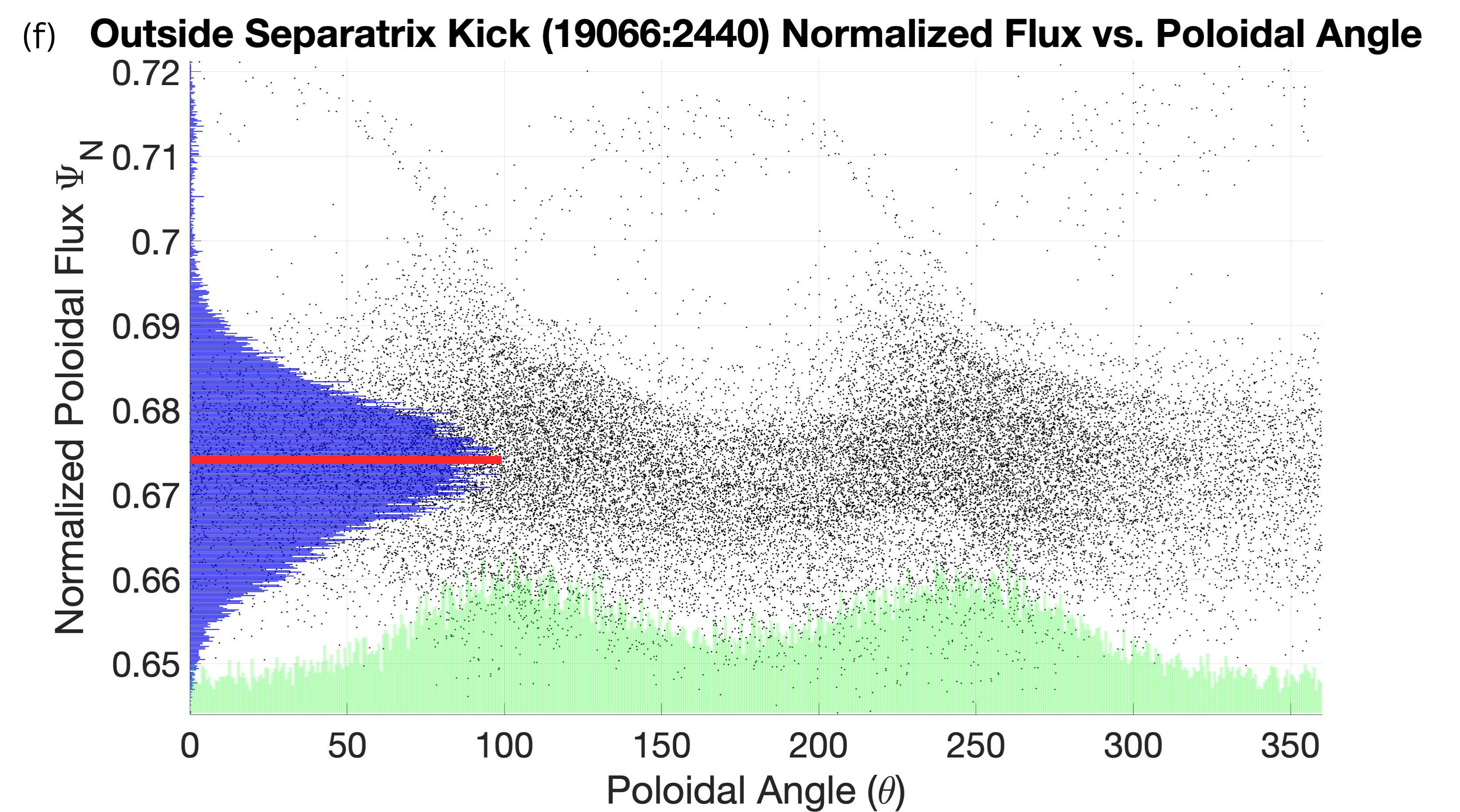}
  \caption{Histograms of final position and final angle location overlayed on final Poincare plot for each kick location, O-point (\textit{a,b}), X-point (\textit{c,d}), and location outside the separatrix (\textit{e,f}), tested with shot 196099 at time slices 2400 (left column) and 2440 ms (right column). Red lines indicate the launch position.}
\label{fig:normpolhist}
\end{figure}

Figure \ref{fig:normpolhist} shows final position vs. final poloidal angle location overlayed on the final Poincare plot for each launch location (O-point, X-point, and outside the separatrix) for shot 196099 at time slices $2400ms$ ($2/1$-dominated structure) and $2440 ms$ ($4/2$-dominated structure). These plots provide a visualization of radial and angular spread for the launched tracers with respect to the island structures.

In the case of the tracers launched from the O-points (fig. \ref{fig:normpolhist} a,d), we see that tracer electrons launched from the O-point tend to remain closer to the O-points. This is evident from the angular distributions being centered around poloidal angles corresponding to the island's O-point locations in each case. However, for the $4/2$ case in Figure \ref{fig:normpolhist} d), two small peaks arise, representing the populations of tracers that become trapped in the emerging second O-point of the island.

When launched from the X-point (fig. \ref{fig:normpolhist} b,e), tracer electrons in the $4/2$-dominated case have a more peaked final angular distribution than in the $2/1$ case. This result suggests that, in the $4/2$ case, tracer electrons remain more localized around the X-points than before in the $2/1$ case. This may be due to the emergence of a smaller X-point within the separatrix of the larger island (see fig. \ref{fig:196099_vacuum} b).

for both the $2/1$ and $4/2$ cases, tracers launched from outside the separatrix (fig. \ref{fig:normpolhist} c,f) exhibit similar angular distributions centered around the island's X-points. This confirms the observation that tracers launched from outside the separatrix mostly diffuse through the island X-points.

\subsection{Quantifying Stochasticity}
Figure \ref{fig:196099_icfpwidthviow} showed that, as the $q=2/1$ island rotates and bifurcates into a $q=4/2$ island, the width of the stochastic region in the edge plasma decreased to its minimum. At the time when the $q=2/1$ island dominates ($2400 ms$), the stochastic region at the edge is wider. Further quantification of these changes in stochastic regions and the impact of the stochastic layer in the edge is analyzed by constructing histograms of magnetic field line displacements from TRIP3D simulations of the vacuum field topology at times before and after the bifurcation.

\begin{figure}
  \centering   
  \includegraphics[width = .45\linewidth]{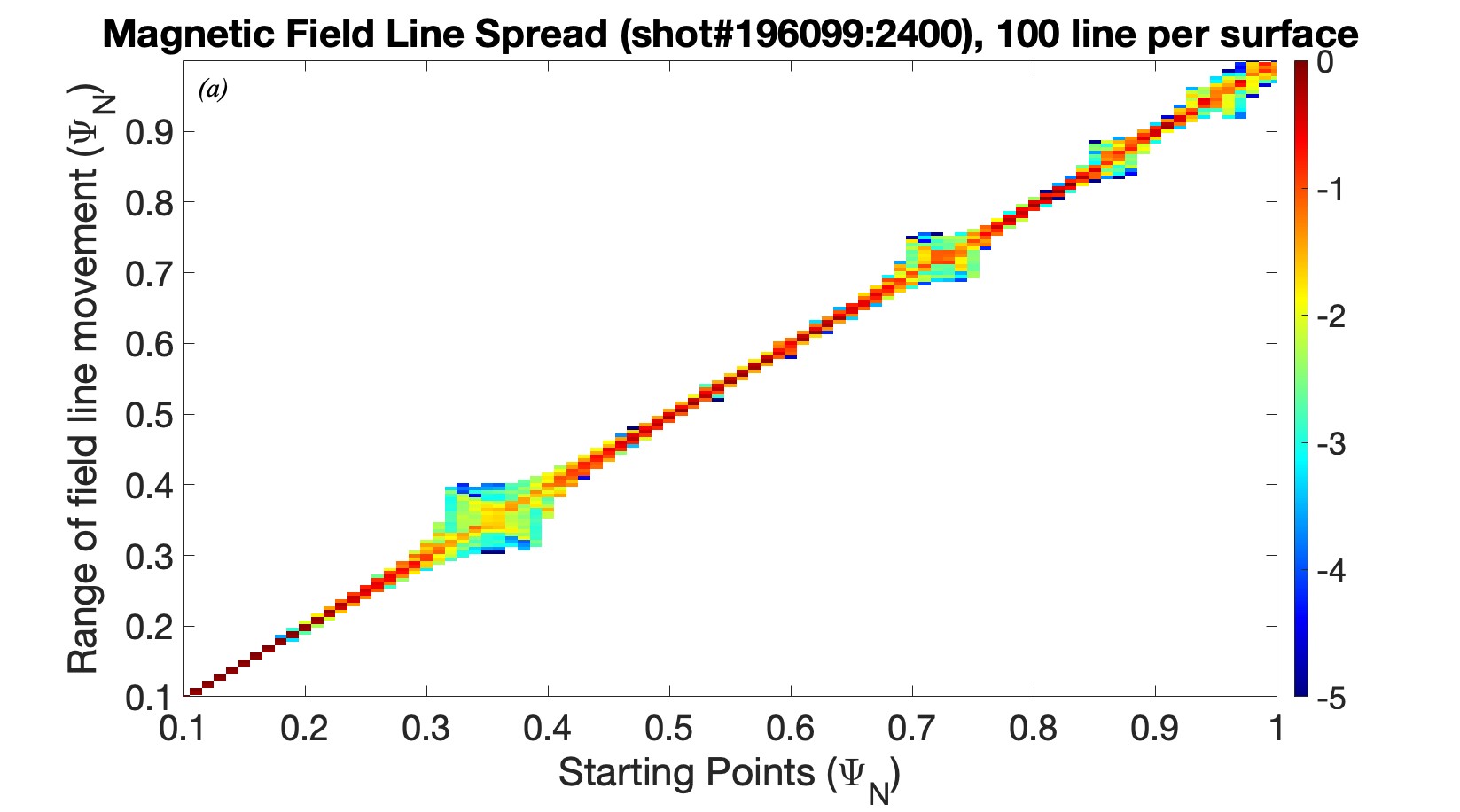}
  \includegraphics[width = .45\linewidth]{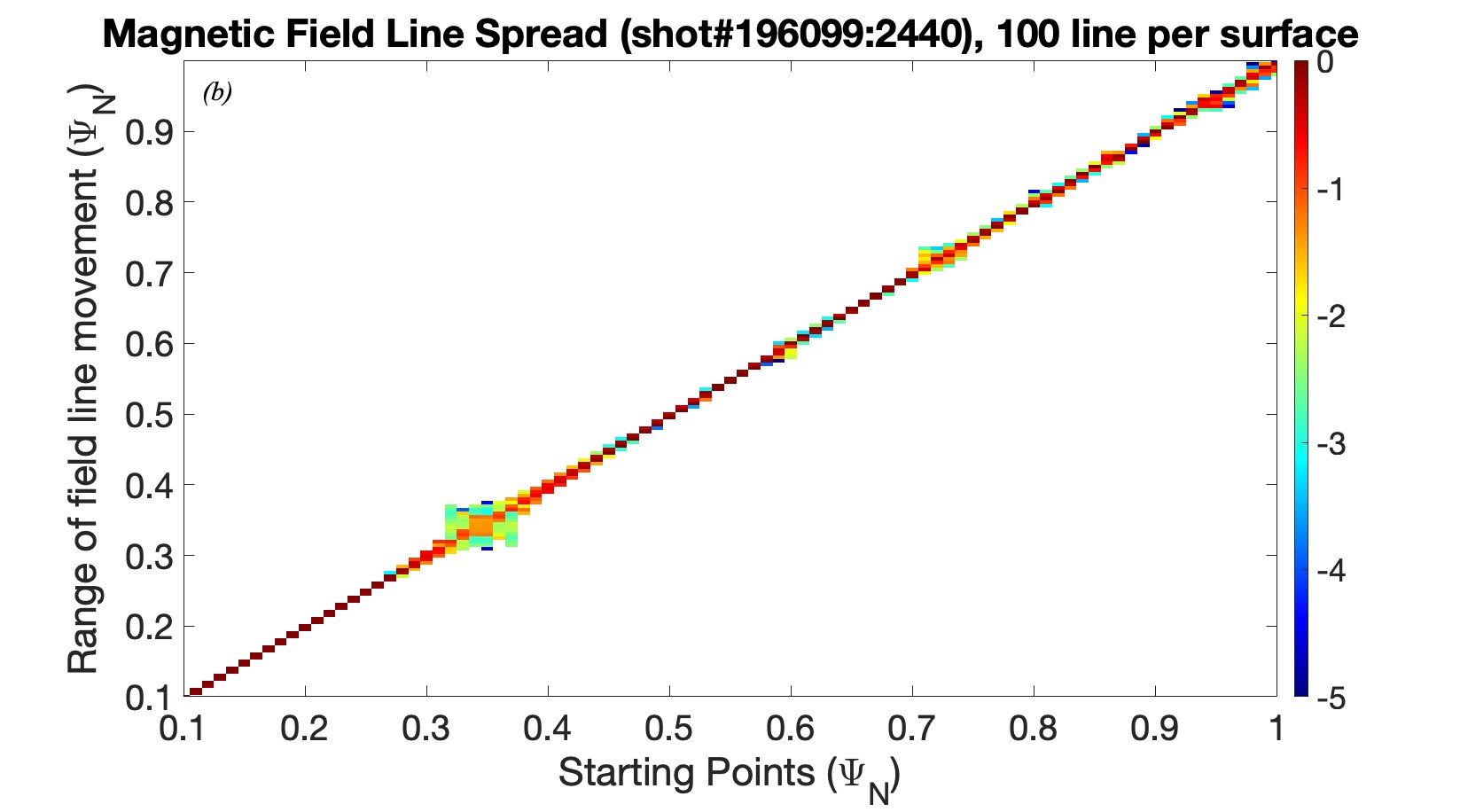}
  \caption{TRIP3D histograms of Magnetic field line spread for shot 196099 
 at (a) 2400 ms and (b) 2440 ms. Here, $\Psi_{N}$ is the range of field line movement from their original position.}
\label{fig:blinespreadhist}
\end{figure}

Figure \ref{fig:blinespreadhist} shows histograms of magnetic field line displacements from their original position. Here, the color bar represents a log scale of the normalized number of times a B-line intersects a given location in space. Red regions represent nested magnetic surfaces or locations in space where the lines did not deviate significantly from their initial positions as the simulation advanced. Broad, rhombus-shaped regions on the plots indicate the locations of magnetic island chains, and irregular, blue regions suggest the stochastization of the field lines. Figure \ref{fig:blinespreadhist} a) shows that, before the bifurcation, regions of stochasticity emerge around the island structures as indicated by the blue/green edges of the rhombus shapes at the location of the islands at rational surfaces. When the $q=4/2$ mode dominates (fig. \ref{fig:blinespreadhist} b), stochasticity still exists around the island structures; however, the range of field line movement is much smaller. Island width decreases during the bifurcation, resulting in a smaller range of field line diffusion. While the case where the $q=2/1$ mode dominates does exhibit a larger stochastic edge than in the $q=4/2$ case, neither feature is significant. Thus, we expect that the energetic electrons detected in the experiment likely results from stochasticization of core island surfaces rather than stochastic edge effects.

\begin{figure}
  \centering   
  \includegraphics[width = .45\linewidth]{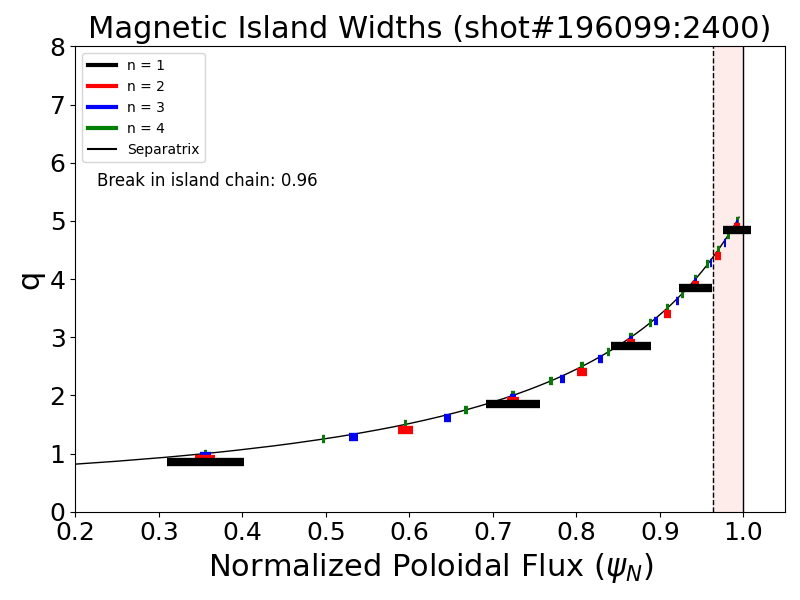}
  \includegraphics[width = .45\linewidth]{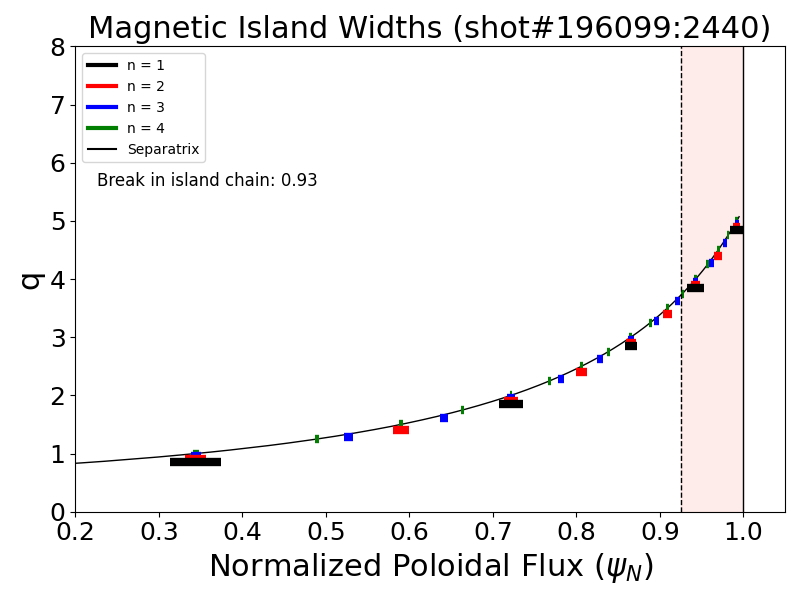}
  \caption{SURFMN island widths
 at (a) 2400 ms and (b) 2440 ms.}
\label{fig:surfmnisles}
\end{figure}

The widths of all islands created by the $n=1$ perturbation are calculated using SURFMN \cite{Schaffer_surfmn}. SURFMN calculates the Fourier harmonics of the magnetic field on flux surfaces to identify resonant modes and magnetic perturbations. Figure \ref{fig:surfmnisles} shows how island width decreases during the bifurcation. It should be noted that a reduction of island widths across all rational surfaces is seen at the time where the $q=4/2$ mode dominates the $q=2$ surface. Further, a reduction in the edge island overlap also occurs during the bifurcation. This island overlap is a transition point where magnetic surfaces are no longer continuous nested flux surfaces or coherent magnetic islands, but rather become stochastic or disconnected.

\begin{figure}
  \centering   
  \includegraphics[width = .45\linewidth]{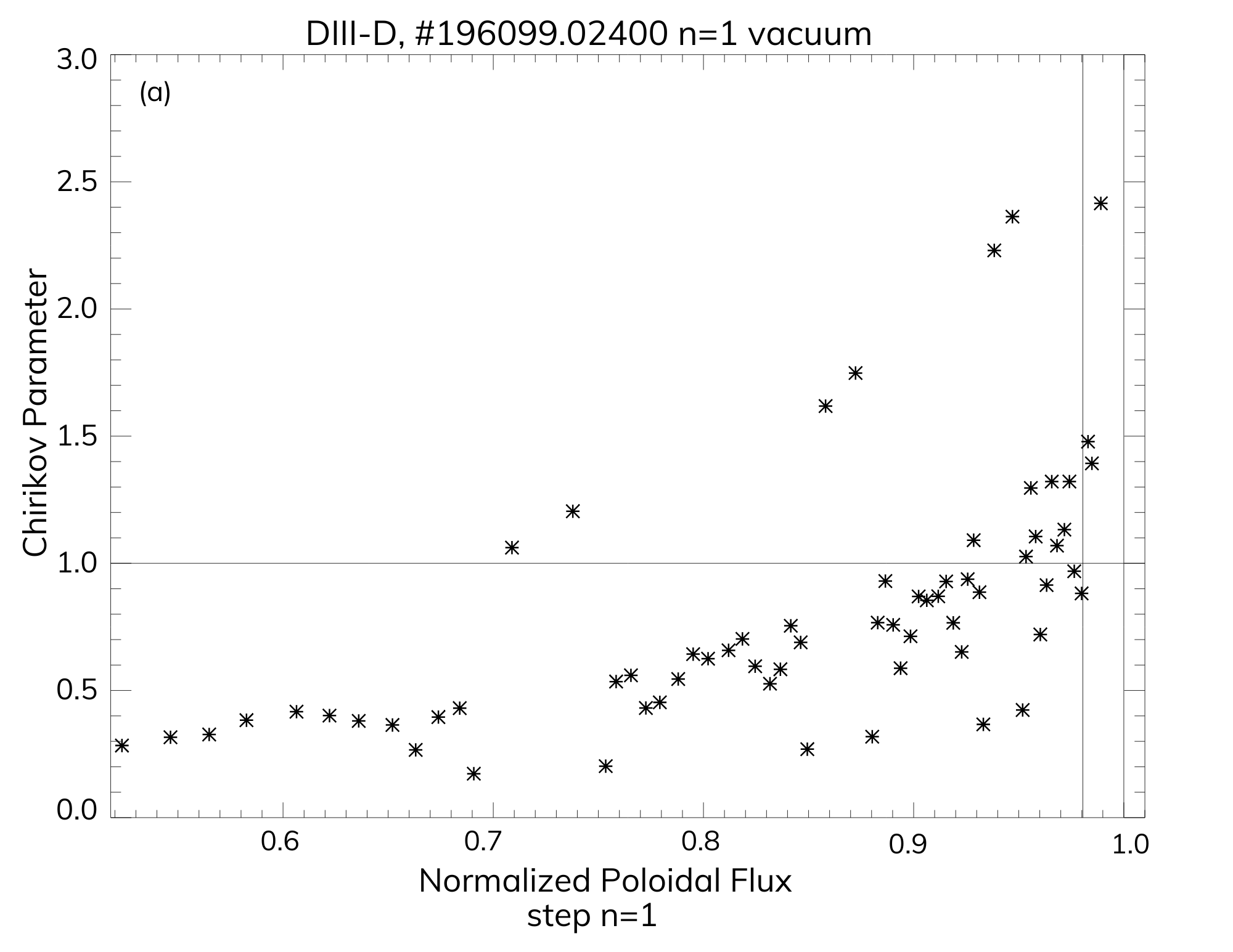}
  \includegraphics[width = .45\linewidth]{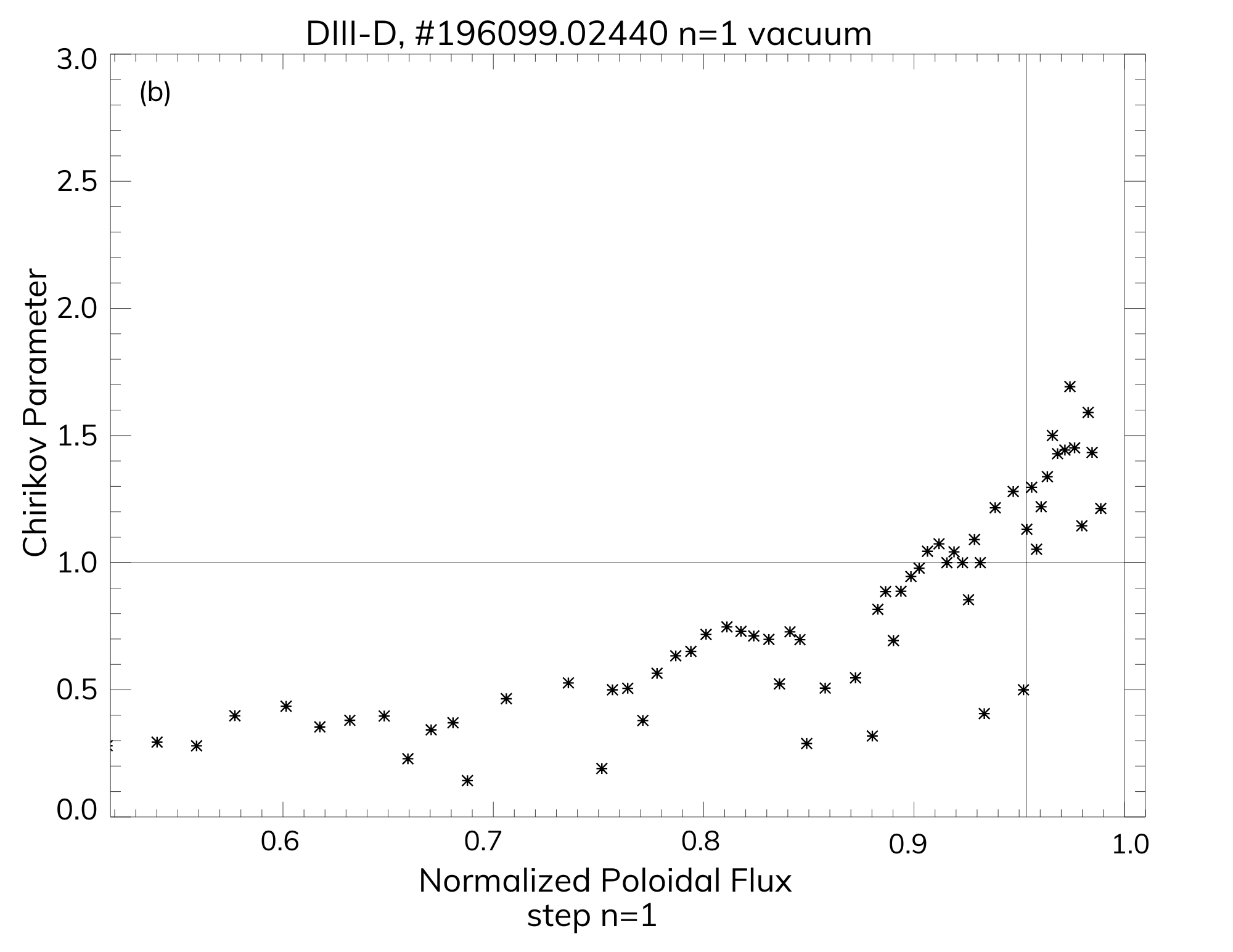}
  \caption{SURFMN calculation of the Chirikov parameter
 at (a) 2400 ms and (b) 2440 ms.}
\label{fig:chirikov}
\end{figure}

Using SURFMN, the Chirikov parameter is also calculated. The Chirikov parameter characterizes the ergodic structure due to overlapping islands and is commonly used as a signature of chaaotic or stochastic field lines \cite{chirikov1979}, \cite{chirikov}. This parameter is calculated by taking the ratio of the separation between two adjacent resonances and the width of a single resonance. When this value exceeds 1, neighboring resonances overlap significantly, suggesting chaotic motion. Figure \ref{fig:chirikov} shows the results of the Chirikov parameter calculation for times before and after the island bifurcation for the region $\psi_N>0.5$ to highlight the location of the $q=2$ surface. At the time when the $2/1$ mode is dominant, we see the Chirikov parameter greater than 1 at the location of the $n=1$ perturbations (with q=2 islands located around $\psi_N=0.72$). When the $4/2$ island is dominant, the Chirikov parameter at those locations is reduced, remaining under 1 for all core surfaces until around $\psi_N>0.9$. This suggests that the I-coil manipulation used for rotation of the island on the $q=2$ surface cased an efficient reduction of the $n=1$ perturbation at all surfaces. This result further supports the observation that the large $2/1$ island exhibit stochasticity inside the separatrix, while the $4/2$ island does not, as seen in Figure \ref{fig:196099_vacuum}. Thus we can expect that in cases where large (or wide) islands dominate the magnetic field topology, stochasticity arises within island separatrices, which can alter the local electron diffusion.

\section{Discussion}\label{sec:discuss}

The present study demonstrates that the topology of magnetic islands strongly influences electron diffusive transport. O-points act as effective traps, confining electrons within islands and leading to subdiffusive behavior, particularly in larger islands such as the $2/1$ case. In contrast, X-points serve as channels for efficient transport across the island, facilitating radial spreading of electrons, with the effect increasing with the size and number of X-points. Larger X-points observed in the 2/1 island are associated with even more efficient transport, highlighting the critical role of island geometry in governing electron motion. Island bifurcation, such as the transition from a $2/1$-dominated to a $4/2$-dominated structure, generates additional X-points and substructure that results in asymmetric and superdiffusive behavior. 

These findings are consistent with previous use of TRIP3D with a collisional operator in NSTX-U simulations \cite{wu_topological_2019}, where test electrons launched within islands preferentially remained confined over finite times, producing higher densities inside the islands compared to the surrounding plasma. The subdiffusive trapping at O-points observed here confirms the island confinement effect reported in those simulations. At the same time, our observations of enhanced transport at X-points align with prior histogram analyses, which show radial spreading along separatrices. 

Additionally, theoretical work \cite{boozer2016} has suggested that island width is related to the critical energy of confined electrons. This is consistent with our observation that the wider $2/1$ island is more efficient in trapping electrons launched from the O-points as compared to the narrower $4/2$ case. We build on this by suggesting that the bifurcation into a narrower island with emerging substructure can further change the electron diffusion from classical to superdiffusive, thus resulting in electron detrapping. 

Understanding electron diffusion in different island topologies provides a further foundation for understanding what additional mechanisms can act to accelerate electrons to relativistic energies. The study by Boozer \cite{boozer2016} proposed that electrons trapped in islands during plasma disruption will be accelerated to relativistic energies due to the thermal quench and the observation that island surfaces do not stochasticize as easily a other surfaces. Our study finds that wider islands will be more efficient in trapping electrons due to the subdiffusion observed around the O-points. However, we showed that the controlled island bifurcation leads to a crossover from subdiffusion to classical diffusion and event superdiffusion, thus facilitating island detrapping. We further observed that the I-coil rotation used in our experiments causes suppression of the island widths at each $n=1$ rational surface, which may be an efficient mechanism for preventing electron trapping in islands during disruptions. 

 The dependence of the electron diffusive transport on the island topology can inform strategies for controlling electron confinement and losses, which is particularly important for predicting and managing energetic electrons. Further understanding how the number and size of X-points and the island bifurcation at different surfaces influence diffusive transport can guide the design of magnetic perturbations to control electron escape before disruptions occur. More broadly, these results provide a framework for predicting how changes in the electron diffusion can enable other mechanisms causing the generation of energetic electron populations in tokamaks and stellarators.

\section{Conclusion}\label{sec:conclude}
Frontiers experiments at the DIII-D tokamak demonstrated controlled repeated bifurcation of the $2/1$ island into a $4/2$ structure using I-coil perturbations. Observations from hard X-ray detectors suggested that the changing magnetic field topology coincides with bursts of energetic electrons. Motivated by these experiments, the present study explored the fundamental question of the relationship between electron diffusion and changing magnetic field topology, specifically the effect of island bifurcation. Analysis through the implementation of a collision operator in the magnetic field line tracing code TRIP3D provided visualization of electron diffusion when tracer electrons are launched near O-points, X-points, or outside the separatrix of a $2/1$ and a $4/2$-dominated island structure. Distributions of final tracer positions provided statistical evidence that the electron diffusion regime is affected by the change in magnetic topology. It was found that (i) electrons exhibit subdiffusion, or trapping, around the O-point of the wide $2/1$ island, (ii) the emergence of sub-structure in the $4/2$ case causes asymmetric diffusion and crossover to superdiffusion, (iii) the radial extent of island X-points acts as an efficient mechanism for transport of electrons across the islands. 

Histograms of magnetic field line deviations from starting positions and calculation of the Chirikov parameter were used to quantify stochasticity across multiple surfaces from core to edge plasma. It was found that stochasticity is enhanced around the locations of wide islands before the bifurcation, and suppressed by the bifurcation due to narrow island widths at the same location. Future work is needed to model X-point tangles to investigate how X-point stochasticization changes electron diffusion in a way that facilitates electron detrapping from magnetic islands. Additional experiments can focus on characterizing island bifurcation at different rational surfaces with the goal of determining whether controlled bifurcation can serve as a method for controlled electron release. Significantly, the experimental methods demonstrated here rely only on external coils and scintillators—tools that are expected to be present in future fusion power plants (FPPs), which will be constrained in diagnostic capabilities. This makes the approach not only practical but scalable. Controlled bifurcation, measured via these minimal tools, may offer a simple, effective disruption mitigation strategy in reactor-scale devices. Expanding these experiments across a range of plasma scenarios and facilities will help validate and generalize these findings for application in future fusion power plants.

%

\ack{The authors thank the DIII-D team for their support of the experiments and diagnostics that made this work possible.}

\funding{This report was prepared as an account of work sponsored by an agency of the United States Government. Neither the United States Government nor any agency thereof, nor any of their employees, makes any warranty, express or implied, or assumes any legal liability or responsibility for the accuracy, completeness, or usefulness of any information, apparatus, product, or process disclosed, or represents that its use would not infringe privately owned rights. Reference herein to any specific commercial product, process, or service by trade name, trademark, manufacturer, or otherwise does not necessarily constitute or imply its endorsement, recommendation, or favoring by the United States Government or any agency thereof. The views and opinions of authors expressed herein do not necessarily state or reflect those of the United States Government or any agency thereof.\\
This material is based upon work supported by the U.S. Department of Energy, Office of Science, Office of Fusion Energy Sciences, using the DIII-D National Fusion Facility, a DOE Office of Science user facility, under Award(s) DE-FC02-04ER54698, DE-FG02-05ER54809, DE-SC0023476, DE-SC0023061, DE-SC0021185, DE-FG02-97ER54415, DE-FG02-07ER54917,  DE-SC0022969, DE-FG02-97ER54415 and DE-SC0023061. This work was supported by NSF-PHY-2440328. Part of the data analysis was performed using the OMFIT integrated modeling framework \cite{OMFIT2015}. Work supported by US DoE under the Science Undergraduate Laboratory Internship (SULI) program.}

\roles{J. Eskew: Conceptualization, Methodology, Formal analysis, Investigation, Writing – original draft.\\
D. M. Orlov: Methodology, Supervision, Writing – review and editing.\\
B. Andrew: Investigation.\\
E. Bursch: Investigation.\\
M. Koepke: Methodology.\\
F. Skiff: Methodology.\\
M. E. Austin: Methodology.\\
T. Cote: Methodology, Investigation, Writing – review and editing.\\
C. Marini: Methodology, Investigation.\\
E. G. Kostadinova: Methodology, Supervision, Writing – review and editing.\\

}

\data{Data supporting the findings of this study, including processed magnetic field line and tracer particle outputs from the TRIP3D simulations, will be made available from the corresponding author upon reasonable request.}


\bibliographystyle{plain}

\end{document}